\documentclass[preprint2]{proto}
\usepackage{times}
\usepackage{xcolor}
\usepackage{amsmath}

\begin{document}

\title{\textbf{\LARGE OB Associations }}

\author {\textbf{\large Nicholas J. Wright}}
\affil{\small\em Astrophysics Group, Keele University, Keele ST5 5BG, UK}
\author {\textbf{\large Simon Goodwin}}
\affil{\small\em Department of Physics and Astronomy, University of Sheffield, Sheffield, S3 7RH, UK}
\author {\textbf{\large Robin D. Jeffries}}
\affil{\small\em Astrophysics Group, Keele University, Keele ST5 5BG, UK}
\author {\textbf{\large Marina Kounkel}}
\affil{\small\em Department of Physics and Astronomy, Vanderbilt University, VU Station 1807, Nashville, TN 37235, USA}
\author {\textbf{\large Eleonora Zari}}
\affil{\small\em Max-Planck-Institut f\"ur Astronomie, K\"onigstuhl 17 D-69117 Heidelberg, Germany }

\begin{abstract}
\baselineskip = 11pt
\leftskip = 0.65in 
\rightskip = 0.65in
\parindent=1pc
{\small

OB associations are low-density groups of young stars that are dispersing from their birth environment into the Galactic field. They are important for understanding the star formation process, early stellar evolution, the properties and distribution of young stars and the processes by which young stellar groups disperse. Recent observations, particularly from {\it Gaia}, have shown that associations are highly complex, with a high degree of spatial, kinematic and temporal substructure. The kinematics of associations have shown them to be globally unbound and expanding, with the majority of recent studies revealing evidence for clear expansion patterns in the association subgroups, suggesting the subgroups were more compact in the past. This expansion is often non-isotropic, arguing against a simple explosive expansion, as predicted by some models of residual gas expulsion. The star formation histories of associations are often complex, exhibit moderate age spreads and temporal substructure, but so far have failed to reveal simple patterns of star formation propagation (e.g., triggering). These results have challenged the historical paradigm of the origin of associations as the expanded remnants of dense star clusters and suggests instead that they originate as highly substructured systems without a linear star formation history, but with multiple clumps of stars that have since expanded and begun to overlap, producing the complex systems we observe today. This has wide-ranging consequences for the early formation environments of most stars and planetary systems, including our own Solar System.

~\\~\\~\\~}
 
\end{abstract}

\section{\textbf{Introduction}}

OB associations are low-density groups of young stars, typically containing many prominent OB stars as well as numerous low-mass stars \citep{blaauw1991,brown1999}. They are technically divided into OB associations, which contain bright OB stars, and T associations, containing prominent T-Tauri stars, though other than their total mass there are no other differences and the term {\it association} is often used to refer to both types.

The low space densities ($< 0.1$~M$_\odot$ pc$^{-3}$) of associations make them dynamically unstable to Galactic tidal forces and therefore over time they should disperse. The fact that they exhibit some spatial and kinematic concentration (despite most likely being unbound) and contain short-lived OB stars implies that they must be young, and are therefore valuable tracers of the star formation process, allowing us to study the propagation of star formation and the role of feedback in triggering or halting star formation.

Historically, OB associations were vital for identifying groups of OB stars and calibrating their luminosity scale \citep{morgan1953,humphreys1978}. This work led to the first census and catalogue of classical OB associations by \citet{ruprecht1966}, which was compiled and standardised from earlier works. This catalogue was most-recently updated by \citet{wright2020}, though many systems still remain poorly defined and studied. More recently, associations have provided large samples of unobscured young stars that are useful for studies of the initial mass function, the frequency and properties of multiple systems, protoplanetary disks and planetary systems \citep{massey1995,kouwenhoven2007,kalas2015}.

The origin of associations is still debated; according to some they are the short-lived expanded remnants of dense star clusters \citep{kroupa2011}, while others suggest they are the result of multiple, sub-structured star formation events occurring over large spatial scales or at low-density \citep{miller1978,kruijssen2012}.

\subsection{Overview of association properties}

Associations have total stellar masses between a few hundred and a few tens of thousand of solar masses. Their dimensions range from a few tens to a hundred parsecs (Section~\ref{sec:structure}), giving them typical densities of 0.001 -- 0.1 M$_\odot$ pc$^{-3}$. They are highly asymmetric and substructured (Section~\ref{sec:structure}), and often contain open clusters or star forming regions within their boundaries. Associations have 1D velocity dispersions of a few km~s$^{-1}$ and are generally expected to be gravitationally unbound (Section~\ref{sec:kinematics}) and expanding (Section~\ref{sec:expansion}).

The ages of associations range from a few to a few tens of Myr, with both considerable age spreads (comparable to the age of the system) and resolvable age substructure (Sections \ref{sec:ages} and \ref{sec:kinages}). The lower age limit historically separated associations from obscured and embedded systems \citep[embedded clusters,][]{lada2003}, while the upper limit comes from the difficulty identifying low-density groups of stars older than this (Section~\ref{sec:identifying}) -- though see \citet{kounkel2019} for an attempt to extend this to older systems.

Associations are observed throughout the disk of our galaxy and their distribution has been used to trace the spiral structure of the Milky Way (Section~\ref{sec:distribution}), map previous generations of young stars and study the propagation of star formation (Section~\ref{sec:propagation}).

\subsection{On the definition of OB associations}

OB associations were originally defined as systems of stars that are very young compared to the age of the galaxy, have a common origin, and a stellar density lower than the Galactic field \citep{ambartsumian1947,ambartsumian1949}. This definition, particularly the last part, distinguished associations from the other types of stellar group known at the time, open and globular clusters. Similar definitions were employed in the landmark studies by \citet{blaauw1964}, \citet{garmany1994}, \citet{dezeeuw1999} and \citet{lada2003}.

With the discovery of very young {\it embedded clusters} some studies argued that most embedded clusters and young star forming regions evolve into associations \citep[e.g.,][]{lada2003,zinnecker2007}. In this context \citet{lada2003} separated star clusters into embedded and open clusters depending on their association with interstellar matter. They argued that as embedded clusters emerge from their natal cloud they either survive as gravitationally bound open clusters or expand as unbound associations. From this a more physical definition of associations emerged, one in which associations are specifically gravitationally unbound \citep[e.g.,][]{gouliermis2018,wright2020}, a definition in harmony with that of a star cluster as being gravitationally bound \citep[e.g.,][]{portegieszwart2010}. 

The difficulty with this physical definition is that, even for relatively isolated clusters and associations, it can be difficult to reliably determine whether a group of stars is gravitationally bound. This is not usually a problem for most classical OB associations since their stellar density is sufficiently low that there is no doubt that they are unbound. However, it can be an issue for particularly young systems or where interstellar matter is present, since the precise mass and spatial distribution of this matter is much harder to determine than it is for stars. Furthermore, recent studies suggest star formation occurs over a continuum of densities rather than at either low or high densities \citep[e.g.,][]{bressert2010,kruijssen2012,kerr2021} and therefore applying a density threshold to very young (still embedded) systems would be meaningless.

For this reason we choose to define associations as {\it groups of young stars with a stellar density lower than that of the Galactic field and that are not strongly associated with interstellar matter}. The density definition separates associations from open clusters, while the interstellar matter definition separates them from star forming regions and embedded clusters. This definition is advantageous because it is both purely observational and preserves the historical definition of associations.

One note to this definition is that since associations are large and highly substructured they may (and often do) contain open or embedded clusters and star forming regions within their borders. The presence of star clusters as the nuclei of associations has been known since their original definition \citep{ambartsumian1949}, while the the existence of star forming regions within their borders is evident from many of the nearest associations such as Sco-Cen or Orion OB1. These denser or younger regions are clearly part of a larger association, though they would not be classified as associations in their own right.

\subsection{This and past reviews}

This review summarises recent results on associations and our current knowledge of their properties and origins. It builds on a strong history of review articles on this topic dating from both the pre-{\it Hipparcos} period \citep{blaauw1964,blaauw1991}, the {\it Hipparcos} era \citep[during which many great discoveries concering associations were made, e.g.,][]{brown1999} and the early {\it Gaia} era \citep{wright2020}. The wealth of studies concerning associations that have emerged over the last few years thanks to {\it Gaia} data have accelerated changes in our understanding of these systems, making this review very timely.

\section{\textbf{Identifying members of associations}}
\label{sec:identifying}

Historically, the main obstacle to studying associations was the difficulty reliably identifying their members, particularly the more-numerous low-mass stars that cannot easily be distinguished from older field stars. In this section we will discuss how the young low- and high-mass members of associations can be identified, the limitations of these techniques, the methods used to verify youth using spectroscopy, and how samples of young stars are commonly divided into different associations or subgroups.

\subsection{Identifying young stars}\label{sec:identifying}

There are numerous methods that can be used to identify candidate young stars, depending on their effective temperature and age. Each of these methods requires different data, is effective under different circumstances, and has certain biases that can limit the effectiveness of the method.

Massive members of associations can be identified using photometry. Following the reddening-invariant colour method pioneered by \cite{johnson1953}, which maps out where stars lie in a colour-colour diagram depending on their reddening, \cite{mohr-smith2015} selected massive members of the cluster Westerlund 2 and surrounding associations using the $(u-g, g-r)$ colour-colour diagram. Such diagrams show that reddened OB stars are located above the main stellar locus, and thus can easily be separated from the older field star population. Follow-up spectroscopy has demonstrated that this method produces levels of contamination from late-type stars as low as 3\% \citep{mohr-smith2017}. In regions of high extinction (where $u$-band photometry is unavailable) the optical identification of OB stars becomes difficult. It is therefore necessary to employ either near-IR or a combination of optical and near-IR photometry. \cite{comeron2002} used $JHK_s$ 2MASS photometry to select massive stars in Cygnus OB2, while \cite{poggio2018}, \cite{zari2021} and \citet{quintana2021} used a combination of optical (\textit{Gaia}) and IR (2MASS) photometry to identify massive stars in the Galaxy within 4-5 kpc from the Sun. These samples are  mainly contaminated by evolved massive or intermediate mass stars (e.g. yellow and blue super-giants), but they are substantially free from RGB and AGB stars \citep[see Section 2.3 in][]{zari2021}. Finally, late B-type stars can live for $\approx$100 Myr, and can therefore be much older than the typical ages of associations (see Section \ref{sec:ages}); to definitely assign such stars to single associations it is necessary to combine photometric identification with other methods, such as kinematics.

Historically, low-mass stars with ages of a few Myr have been difficult to identify, however dusty young stellar objects (such as protostars or pre-main sequence stars with disks) can be selected through near- and mid-IR photometry, allowing for their identification with targeted surveys of individual star forming regions with telescopes such as Spitzer or Herschel \citep{gutermuth2009,fischer2017,winston2020}, or through all-sky surveys such as AKARI or WISE \citep{toth2014,marton2016}. These methods are effective, but heavily biased towards very young stars ($<5$ Myr) that retain their disks, while disk-free candidate YSOs identified using IR photometry appear to be mainly contaminants \citep{manara2018}. Identifying older stars and the substantial fraction of young stars without disks requires alternative methods.

\begin{figure}[h]
    \centering
    \includegraphics[width = \hsize]{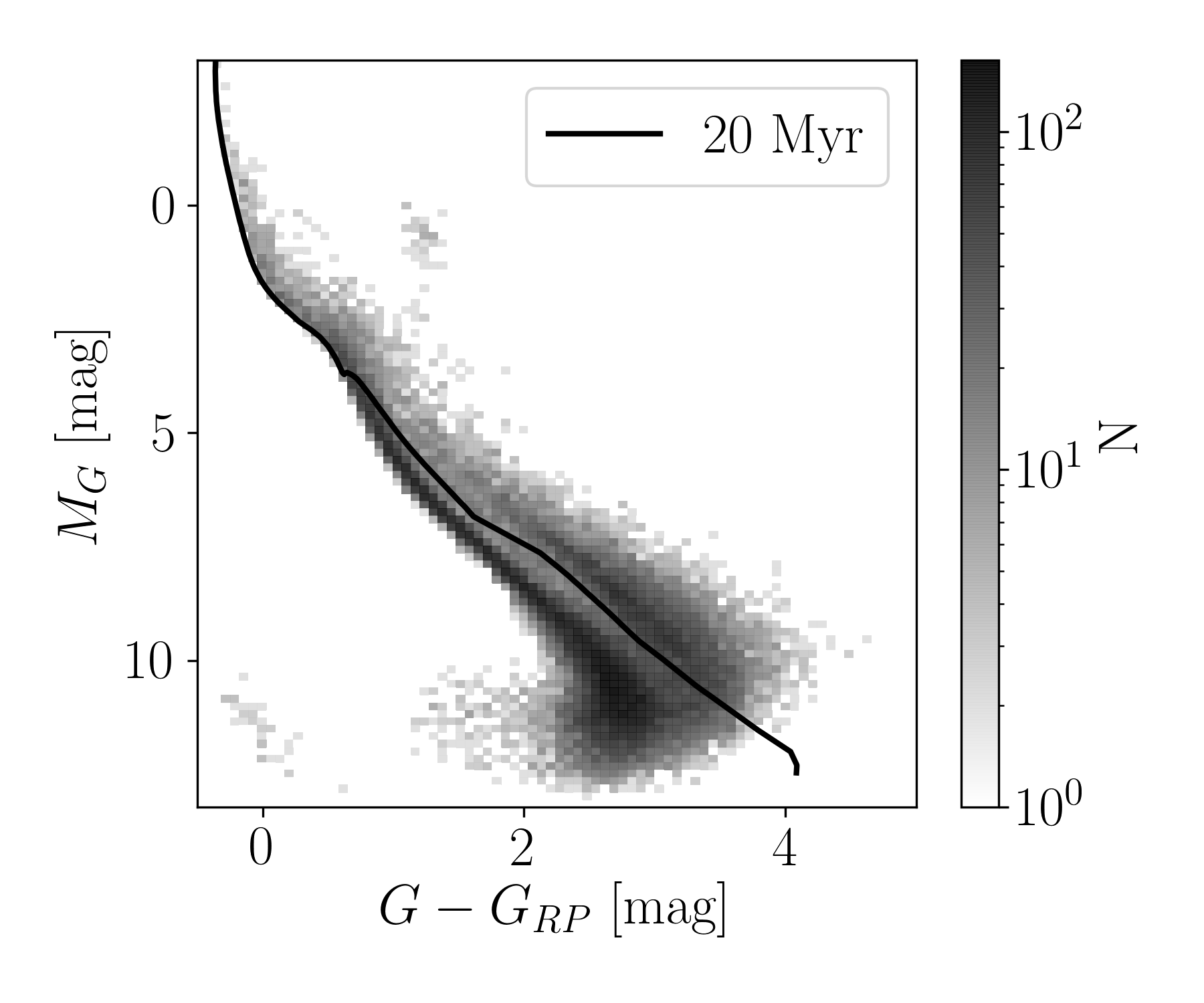}
    \caption{Colour-absolute magnitude diagram for stars in Orion based on \textit{Gaia} EDR3. The black solid line is a 20 Myr solar metallicity PARSEC isochrone \citep{Marigo2017}. For stars with $G-G_{RP} > 1$, where the isochrone separates from the main sequence, isochrones like this can be used to separate the young PMS population above the isochrone from older field stars below it.}
    \label{fig:cmd}
\end{figure}

Most low-mass members of associations are in the pre-main sequence (PMS) phase of stellar evolution. PMS stars are more luminous than main sequence stars of the same $T_{\rm eff}$ and therefore stand out, above the zero-age main sequence, in optical colour-absolute magnitude diagrams (see Fig. \ref{fig:cmd}). This provides an effective and efficient technique to identify young stars that requires only photometry and astrometry. The method is particularly useful for identifying low-mass stars (as low-mass PMS stars are more over-luminous at a given age than more massive stars) and isn't biased towards other properties of the star (e.g., the presence of circumstellar material or magnetic activity), however some care is required when using it. For example, the extinction correction required by this method could introduce selection biases to the sample \citep[see][]{zari2018}. Furthermore, the samples selected may be contaminated by unresolved multiple stars or evolved field stars with uncertain parallaxes, especially for PMS stars at intermediate masses that are less offset from the main sequence than low-mass stars. Finally, as \textit{Gaia} parallax uncertainties increase as a function of magnitude, faint sources at large distances can be excluded when applying cuts on the relative parallax error $\sigma_{\varpi}/\varpi$. Nevertheless, following \textit{Gaia} DR2 this approach has been used in numerous studies to identify PMS stars younger than 20 Myr \citep[e.g.,][]{zari2018,cantat-gaudin2019,damiani2019}. Attempts at identifying PMS stars older than 20 Myr (and younger than 50 Myr) using this method have been made by \cite{kerr2021} and \cite{mcbride2021}, who also propose different methods to estimate and correct for extinction.

In the absence of reliable parallaxes, single epoch optical or near-IR photometry can be combined to identify candidate M-type stars \citep{damiani2018} or employed to determine the size and spatial distribution of the low-mass PMS population of an association \citep[see e.g.][]{sherry2004, bouy2014, zari2017, armstrong2018}. Single epoch photometry alone cannot definitively identify an individual star as a PMS star, thus samples selected based on this technique can include both background giants and foreground main-sequence stars of any age. 

Optical photometric variability is one of the defining characteristics of pre–main-sequence stars \citep{joy1945, herbig1962}. \cite{Briceno2005} used variability combined with position in colour-magnitude diagrams and follow-up spectroscopy to select the low-mass young population in the Orion OB1 association. This method is potentially biased towards PMS stars with only high amplitude photometric variations, nevertheless current and future \textit{Gaia} releases will facilitate all-sky searches for variable PMS stars and the use of variability as an additional selection criterion.

PMS stars are magnetically active, which makes them bright X-ray sources. Studies of low-mass X-ray emitting young stellar populations of associations were carried out using data from the Einstein observatory \citep{walter1994}, and the ROSAT all sky survey \citep{sterzik1995, walter2000}. More recently, the X-ray observatories {\it Chandra} and XMM-{\it Newton} have been very effective at identifying large populations of young, magnetically-active stars, thanks primarily to their arcsecond-scale point spread function and high sensitivity \citep[e.g., the Massive Young star-forming complex Study in Infrared and  X-rays, MYStIX, which studied young stars in 20 Galactic massive star-forming regions,][]{feigelson2013}. However, the small fields of view of these observatories have meant they have preferentially targeted compact star clusters rather than diffuse associations \citep[with the exception of more distant associations, e.g.,][]{wright2010}. All-sky data from eROSITA \citep{merloni2020} will overcome this issue and will allow members of many nearby associations to be identified, as shown by the preliminary study presented in \cite{Schmitt2021} (see \S\ref{sec:futurex-ray}). As for the other techniques presented in this Section, although X-ray selection may weed out a lot of old stars, there will be many false positives remaining. For instance, the spin-down timescales of low-mass stars are $\sim$50 Myr for solar-type stars and longer at lower-masses, so contamination by young field stars is probable. In addition, tidally locked, close binaries can maintain their X-ray activity indefinitely and appear `young' in the HR diagram due to the combined luminosity of the two stars.

In summary there are many different methods for identifying young stars. The advantages and disadvantages of each method, as applied to associations, are a combination of: whether the method is sensitive in the typical 1--50 Myr interval we are interested in; whether they offer any age discrimination within this range; whether they can be applied to isolated stars; and whether the observational inputs are available over wide fields of view. Unfortunately, while each of the methods discussed here allows identification of young stars, they can all introduce contamination by older, unrelated field stars. These methods are therefore each insufficient in isolation to confirm the youth of stars.

\subsection{Confirming the youth of stars}
\label{sec:confirming}

The {\it secure} identification of association members is inextricably linked to confirmation of their age, with the most decisive indicators of youth coming from spectroscopy. Estimates of spectral type or temperature can improve reddening determinations, allowing more accurate placement in the HR diagram (and this is often all that is required to confirm the status of young high-mass stars), but for the more numerous lower mass stars there are more direct methods of determining youth.

\begin{figure}[t]
    \centering
    \includegraphics[width = \hsize]{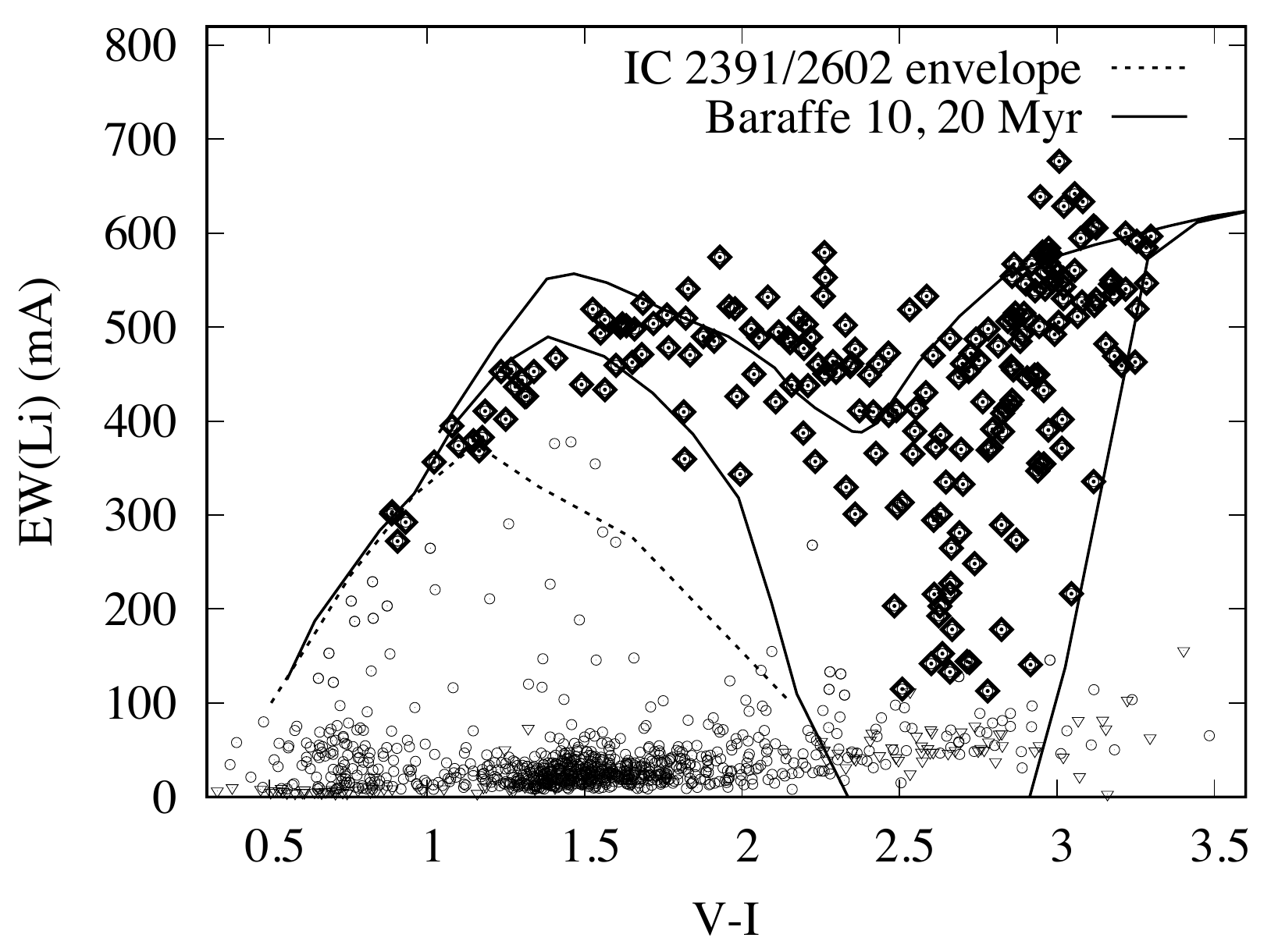}
    \caption{EW of the Li~{\sc i} 6708\AA\ absorption line for stars towards the $\gamma$ Vel cluster and Vela OB2 association. The bold diamonds show objects selected as members of the cluster/association on the basis of their Li abundance and position in the $V$ vs $V-I$ CMD \citep[see][]{jeffries2014}, while the open circles and open triangles (indicating upper limits) are sources not selected as members. The solid lines show theoretical predictions of the strength of this line at ages of 10 (upper) and 20 (lower) Myr \citep{baraffe1998}, while the dashed line indicates an empirical upper limit to the line strength at 50 Myr judged from observations of stars in the IC 2391/2602 clusters. The age discrimination of this Li feature is quite dependent on colour/mass; in particular, it is less sensitive at bluer colours/higher masses and  may not offer a complete selection of members where the association stars are old enough to have depleted their Li in certain mass ranges (e.g. in the mid M-dwarfs in this example).}
    \label{fig:li}
\end{figure}

Lithium is an ephemeral element in the photospheres of low-mass stars; it is consumed by nuclear reactions on the PMS. Whilst a comprehensive understanding of the Li depletion process is still lacking, it is clear that Li abundance can be used as an empirical indicator of youth in a mass-dependent way (see Figure~\ref{fig:li}).  At low masses ($<0.7 M_\odot$) the presence of undepleted Li is a sure sign that a star is younger than about 20 Myr. Conversely, stars which exhibit total Li depletion must be much older, whilst intermediate cases offer some age discrimination \citep{Jeffries2014b}. Li is less diagnostic at higher masses because there is little PMS depletion and depletion on the main sequence is much slower. Using Li abundance as a tracer of young stellar populations in the nearest associations has a long history \citep[e.g.,][]{Martin1998, preibisch1998, Mamajek2002,preibisch2002}. Improvements in instrumentation are now allowing an extension of these techniques to wider fields and greater numbers of stars \citep[e.g.,][]{jeffries2014,Briceno2019, armstrong2018}, however the requirement for reasonably high resolving power and signal-to-noise still limits this technique to associations closer than $\sim 1$~kpc.

Spectroscopic indications of gravity can be used to confirm stellar youth at low masses and can resolve the confusion in absolute colour-magnitude diagrams caused by variability, reddening and binarity. PMS stars have lower gravities than MS stars of similar $T_{\rm eff}$, but for a given PMS age the difference is wider at lower masses, so these techniques are most sensitive for very young stars and/or PMS stars of low-mass ($<1M_\odot$) \citep[see Figure 8 of][for an example]{lopez-valdivia2021}. Surface gravity is difficult to measure directly, although its relative magnitude can be inferred from gravity-sensitive spectral indices in the optical/near IR, such as atomic alkali,  CaH or TiO molecular lines \citep{wilking2005}, empirically constructed spectral indices \citep{damiani2014} or the shape of the $H$-band peak \citep{scholz2009}. Directly-measured surface gravities can be used to calibrate pre-MS evolutionary models \citep{olney2020} and directly infer the age of a low-mass star. In comparison to photometrically derived ages, this method does not underestimate the ages of stars on the binary sequence.

Young stars with a protoplanetary disk are likely to be accreting gas. Classical T Tauri stars can be easily identified through having strong and broad H$\alpha$ emission, even in low-resolution spectra \citep{white2003}. These strong signatures are limited to just a fraction of the low-mass population of an association, reducing from $\sim 60$\% at a 1-2 Myr to $<5$\% at 10\,Myr  \citep{Fedele2010}. The complementary population of ``weak-lined" T-Tauri stars can still be identified from their weak, chromospheric H$\alpha$ emission (along with several other lines, such as Ca II H \& K). However, these relatively weak activity indicators can persist for a few hundred Myr in low-mass stars so such samples inevitably suffer contamination from field stars and active binaries, in a similar way to X-ray-selected stars.

\subsection{Assigning young stars into groups}\label{sec:assigning}

When a sample of candidate young stars has been identified they can be divided into groups based on their distribution in position (plane of the sky and distance) and kinematics (proper motion and radial velocity). This may be performed after the youth of these stars has been verified (see Section~\ref{sec:confirming}) or to confirm youth via the corroboration that a group of stars is spatially or kinematically coherent (the clustering of stars, both spatially and kinematically, decreases as stars age, and therefore can be used as a youth proxy). A combination of kinematics and spectroscopic indicators of youth is increasingly used to get a sample of the low-mass population with minimal contamination \citep[e.g.,][]{Prisinzano2016, Wright2019}.

For compact clusters, stars can be grouped together using just plane-of-the-sky information, often starting with a curated membership list (that removes most of the field contaminants) and then identifying the group using techniques such as the minimum spanning tree \citep{gutermuth2009}, stellar density maps \citep{carpenter2000,megeath2016}, various mixture models \citep{kuhn2014}, or using other plane-of-the-sky clustering techniques \citep{buckner2019}. Some of these approaches require {\it a-priori} assumptions regarding the number of different substructures present in a given population, and therefore the results of such methods should not be used to study the structure of such regions without understanding the biases at work.

For low-density structures such as associations, assigning stars into groups cannot be done using just sky positions. Since associations have relatively small internal velocity dispersions they continue to form coherent structures in velocity space even as they disperse. In the \textit{Hipparcos} era the identification of massive members of associations relied strongly on proper motions \citep[e.g.][]{dezeeuw1999}, sometimes combined with positions and parallaxes \citep{debruijne1999, hoogerwerf1999}, or radial velocities \citep{rizzuto2011}. Multi-epoch spectra are sometimes required to reliably identify binaries (that can introduce discrepant RVs due to the orbital motions), but even without multiple epochs, the broadening of an RV distribution by binaries can be estimated and assigning high membership probabilities based on a single-epoch of spectra is possible for most single and unresolved binary stars \citep[e.g.,][]{jeffries2014,gonzalez2017}, especially in conjunction with other spectroscopic tracers of youth (see Section 2.2).

With distances becoming increasingly common from {\it Gaia}, studies of the clustering and grouping of stars are now performed similarly in a higher-dimensional space (combining position on the sky, parallax, proper motion, and radial velocity when available). As these multi-dimensional data combine different quantities (position and velocity) with different units, some prescription needs to be considered to scale the data to give each dimension an appropriate weight. With higher dimensionality it is possible to select association members without initial identification of candidate young stars (as is commonly needed for 2D clustering). Several clustering algorithms have been used for such a purpose, including DBSCAN \citep{zari2019,liu2021}, HDBSCAN \citep{kounkel2019,kerr2021}, and UPMASK \citep{cantat-gaudin2019}, Gaussian mixture models \citep{kuhn2020}, as well as other custom codes. The typical inputs for these are the minimum number of stars in a group, the typical group dimensions, and the density threshold required to detect groups (though some codes are able to automatically adjust this threshold to detect both compact and extended populations).

In recent years, with a large number of clustering codes and approaches available, as well as the variance that would result in using the same clustering algorithm tuned with different critical parameters, or applied to a differently-selected initial source list, it is common to see different groups identified within the same region of space in different studies \citep[e.g.,][]{kounkel2018,galli2019,zari2019,kos2019,chen2020,krolikowski2021}. With the greater degree of attention given to extended associations (that are not as simple to define as compact star clusters) some confusion may arise regarding which of these overlapping groups should be considered as fundamental. As such, it is important to keep in mind that these algorithms only trace the underlying density distribution of a given population with a greater or lesser sensitivity to the structures at a particular scale, often requiring a trade-off between the ability to identify finer substructure vs the ability to study the population as a whole.

\section{\textbf{Properties of OB associations}}

Associations come in diverse shapes and sizes and can appear, at first sight, a rather heterogeneous group. In this section we review the general properties of associations, with reference to notable individual examples, and the similarities and patterns that exist.

\subsection{Size and internal structure}
\label{sec:structure}

The sizes of associations range from tens to several hundreds of parsecs \citep[e.g.,][]{blaauw1964,blaha1989,gouliermis2018}, though this can vary between studies due to different methods for defining the borders, different membership of the systems included, and on which systems were included within the sample. For example, \citet{garmany1992} measure a mean OB association size of $137 \pm 83$~pc, while \citet{melnik1995} measure an average diameter of $\sim$40~pc. Studies of extra-galactic OB association populations reveal broadly similar size distributions to Galactic associations, depending on the resolution of the data used. For example, \citet{lucke1970} list 122 OB associations in the Large Magellanic Cloud (LMC) with sizes of 15--150~pc and a mean of 80~pc, while \citet{gouliermis2018} used the better-resolved catalogue of associations in the LMC from \citet{bica2008} to measure an average size of 30~pc.

Associations increase in size with age \citep[e.g.,][]{blaauw1964}, as would be expected for gravitationally unbound and expanding systems. Furthermore, associations connected to nebular material are typically smaller \citep{gouliermis2018}. If the association with nebular material is an indication of relative youth, then this also supports a picture whereby associations increase in size as they age.

Associations exhibit considerable internal substructure. This includes the presence of subgroups with different ages and kinematics \citep{blaauw1964,garmany1992}, young or open clusters \citep{ambartsumian1949}, or bright central concentrations \citep{ivanov1987,melnik1995}. In some cases this substructure is very clear, while in other systems it can be revealed using structural diagnostics that quantify physical substructure \citep[e.g.,][]{cartwright2004,wright2014}.

The most well-studied associations, such as Sco-Cen or Orion OB1, have been sub-divided into ``OB subgroups'' historically based on their on-sky distribution \citep[e.g.,][divided Sco-Cen into the subgroups of Upper Sco, Upper Centaurus-Lupus and Lower Centaurus-Crux]{blaauw1964}. This physical substructure is often correlated with kinematic \citep{wright2016,wright2018} or temporal substructure \citep{pecaut2016} that suggest these subgroups are real and not chance over-densities.

Recent studies have combined on-sky positions with parallax, proper motion or radial velocity information to subdivide associations based on their 5- or 6D spatial and kinematic structure \citep[e.g.,][]{kounkel2018,cantat-gaudin2019,berlanas2019}. For example, \cite{damiani2019} and \cite{kerr2021} studied the Sco-Cen association, finding a wealth of sub-groups with different spatial and kinematic properties, which do not adhere well with the classical sub-division of the association. \cite{cantat-gaudin2019} studied the young stars in the Vela OB2 association and found that the region is highly substructured, with seven main kinematic groups overlapping on the sky and whose formation is spread over $\sim$~35~Myr. 

These physical subgroups often show distinct age differences, supporting the correlation between spatial, kinematic and temporal substructure \citep[e.g.,][]{zari2019}. Furthermore, they often extend beyond the bounds of the classically defined OB associations (usually based on the distribution of OB stars) to expose older parts of the association \citep[e.g.,][]{cantat-gaudin2019}.

Many associations also contain open or embedded clusters within their borders that have been considered part of the association. In many cases their ages and kinematics are consistent with being related to the association, such as the $\gamma$~Vel cluster in Vela~OB2 \citep{jeffries2014}, $\rho$~Oph in Sco-Cen \citep{blaauw1991}, or NGC~2353 in CMa OB1 \citep{fitzgerald1990}. In some cases however the open cluster has been shown to either have a very different age or significantly different kinematics, or both, suggesting that the cluster did not form as part of the association and is now just projected against the association \citep[e.g., the NGC~2547 cluster projected against Vela~OB2,][]{sacco2015}.

\subsection{Kinematics, velocity dispersion and virial state}
\label{sec:kinematics}

Kinematic studies of associations were historically limited to their most luminous members and used either {\it Hipparcos} proper motions or radial velocities obtained from individual spectra. The former were of limited precision for detailed kinematic studies (allowing membership to be constrained, but rarely probing internal kinematics), while the latter were scarce and strongly affected by unresolved binarity. It wasn't until the availability of multi-object spectroscopy \citep[e.g.,][]{preibisch2002,furesz2008,da-rio2016} and, later, {\it Gaia} proper motions \citep[e.g.,][]{wright2018,kounkel2018} that detailed kinematic studies of the (more numerous) low-mass members of associations became possible.

From kinematic studies using either radial velocities, proper motions or both, estimates of the velocity dispersion of associations can be calculated. Combining the velocity dispersion with estimates of the total mass of the association, the virial state can be derived.

The accurate calculation of the velocity dispersion from measurement of the radial velocity or proper motion for individual stars requires that the measurement uncertainty and the impact of unresolved binaries be accounted for. The former can be easily quantified, but requires some modelling as the uncertainties are often significantly heteroskedastic or do not follow a normal distribution (e.g., \citealt{jack15} find that the uncertainty distribution of Gaia-ESO Survey radial velocities better follows a Student's t-distribution with extended tails). Unresolved binaries can significantly affect the instantaneous measure of velocity from single-epoch radial velocity surveys, particularly for high-mass stars that are commonly found in binary or multiple systems \citep[e.g.,][]{gieles2010}, and require full modelling \citep[e.g.,][]{cott12}. The impact of unresolved binarity on measured proper motions has not been fully explored, and while smaller than the effect on radial velocities, is still potentially significant \citep{jack20}. Proper motions can also become unusable at large distances due to the distance-dependence of velocities derived from them, a problem that does not affect radial velocities.

The corrected velocity dispersions of associations can show considerable variation, but are typically a few kilometres per second along each axis. For example, \citet{ward2018} studied 18 associations using {\it Gaia} DR1 proper motions and measured velocity dispersions of 3--13~km~s$^{-1}$ with a median of 7~km~s$^{-1}$, and \citet{melnik2020} measured an average velocity dispersion of 4.5~km~s$^{-1}$ from 28 associations studied with {\it Gaia} DR2. Small differences between studies are predominantly due to differences in the association membership lists, many of which date from the 1980s \citep[e.g.,][]{humphreys1978,blaha1989} and need revisiting and updating with modern data \citep[e.g.,][]{quintana2021}.

Immediately apparent from the earliest kinematic studies of associations was that their velocity distributions were far from Gaussian and correlated with spatial position \citep[e.g.,][]{furesz2008,tobin2009}. Later studies using proper motions (that probed more than one kinematic dimension) found that the velocity distributions of associations were also highly anisotropic, with velocity dispersion ratios between the two proper motion axes of at least $\sim$1.5 \citep{wright2016,melnik2017,ward2018} and up to $\sim$6 \citep{melnik2017}. 3D kinematic studies of associations using proper motions and radial velocities are rare, but similar levels of anisotropy have been found \citep[e.g.,][]{wright2018}.

\begin{figure*}[h]
    \centering
    \includegraphics[width = 14cm]{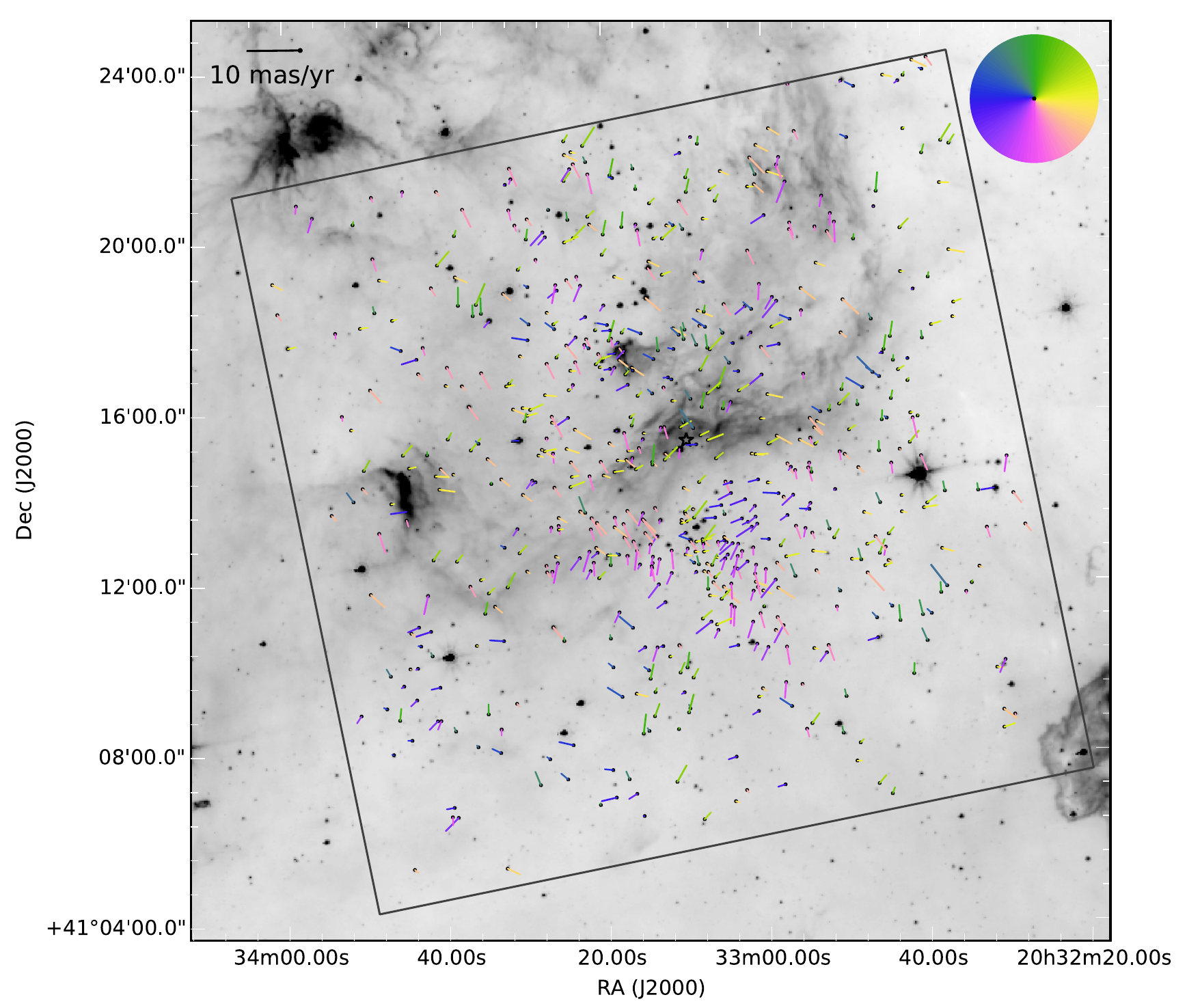}
    \caption{Proper motions for 798 X-ray and spectroscopically selected stars towards Cygnus OB2 (including 16 O-type stars), with kinematic outliers removed. The vectors are coloured based on their direction of motion to highlight kinematic substructure. The grey box shows the border of the X-ray observations used to identify members and the colour wheel in the top-right corner shows the relationship between colour and position angle. Figure from \citet{wright2016}.}
    \label{fig:cygob2_substructure}
\end{figure*}

The anisotropy and non-Gaussianity observed within associations has been attributed to {\it kinematic substructure}, groups of stars in the same area of space with similar kinematics, but within a wider group with more diverse kinematics. An example of this in Cyg OB2 from \citet{wright2016} is shown in Figure~\ref{fig:cygob2_substructure} where the proper motion vectors are coloured according to their position angle on the sky to highlight the kinematic substructure. This kinematic substructure has since been observed in nearby all associations studied \citep[e.g.,][]{wright2018,kounkel2018,cantat-gaudin2019}. There have been many efforts to quantify the kinematic substructure. For example, \citet{arnold2020} analysed the kinematic substructure in Cyg OB2 and found a strong spatial/velocity correlation on sub-pc scales, but no correlations or structures were found on larger scales.

This kinematic substructure represents subgroups of stars within the association, and using sufficiently-accurate kinematic data the subgroups can be separated. There is growing evidence that when association subgroups are identified using kinematics that the velocity dispersion anisotropy of the subgroups is less pronounced \citep[e.g.,][find velocity dispersion ratios of 1.0--1.6 for the subgroups in Vela OB2]{cantat-gaudin2019b}, which may indicate that the true association subgroups have isotropic velocity dispersions or might highlight a bias introduced by using kinematics to identify subgroups.

The virial state of the association can be estimated by combining the measured velocity dispersion with estimates of the total stellar mass of the association (often extrapolated from a subset of the total population and therefore requiring assumptions about the form of the mass function), though see \citet{parker2016} for caveats. Given that associations have a low stellar density compared to gravitationally bound open clusters, it is unsurprising that all associations to date have been found to be super-virial. The ratio between the virial mass and the stellar mass was found by \citet{melnik2017} to range from 10 to 1000, with a median of $\sim$50. Assuming there are no additional forces acting on the members of the association as it expands, then the virial mass will increase as the association expands \citep[since the virial mass is proportional to radius,][]{portegies-zwart2015}. If the expansion is assumed to be linear then the virial mass for an association will increase linearly with time, and as the stellar mass remains approximately constant, the virial to stellar mass ratio will also increase with time.

\subsection{Expansion}
\label{sec:expansion}

Given their low space density, associations should be unbound and in a state of expansion and may have been more compact in the past \citep[e.g.,][]{ambartsumian1947,ambartsumian1949,blaauw1952}. This expansion can be inferred from the observed correlation between radius and density \citep{pfalzner2009}. However, the internal velocity dispersions for many associations are too small to explain their present-day size by expansion from a significantly more compact state \citep[e.g.,][]{preibisch2008,torres2008}, raising questions over this interpretation.

Measuring the expansion of associations is, in principle, very simple. \citet{blaauw1946} put forward the linear expansion model, which assumes that associations had an initially compact configuration and have expanded linearly (in time) from then. For nearby associations it is necessary to account for {\it virtual expansion} caused by the radial motion of the association towards (or away from) the observer, which can cause a false expansion (or contraction) on the sky, even when no physical expansion exists \citep{blaauw1964}. This can be corrected for using radial velocities, either for the bulk motion of the association \citep{brown1997} or for individual stars \citep{wright2018}.

Early kinematic studies struggled to find evidence of expansion in associations. For example, both \citet{wright2016} and \citet{arnold2020} could find no evidence for expansion in Cyg~OB2 from ground-based proper motions, while \citet{ward2018} searched for expansion in 18 OB associations using {\it Gaia} DR1 data, but could not find evidence for expansion in any of their targets. The availability of {\it Gaia} DR2 astrometry lead to expansion being measured in some, but not all, systems. Using {\it Gaia} DR2 data, \citet{melnik2020} studied 28 OB associations and found evidence for expansion within 6 of them. \citet{ward2020} studied the kinematics of 110 OB associations using {\it Gaia} DR2 and argued that their properties were not consistent with expansion from a single, compact configuration, but were more consistent with originating from a highly substructured velocity field.

\begin{figure}
    \centering
    \includegraphics[width = \hsize]{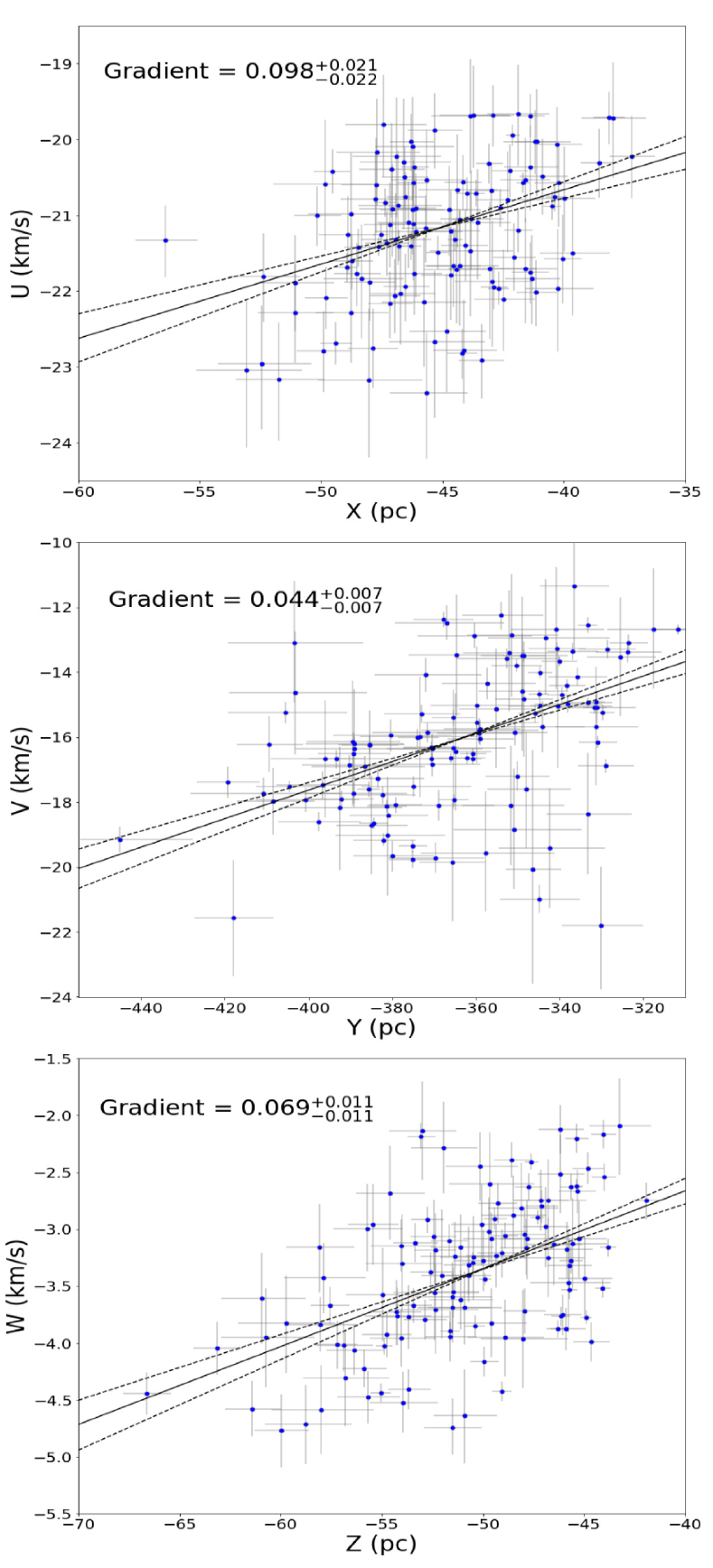}
    \caption{Cartesian position–velocity plots of stars in Vela OB2 with 3D kinematics from \citet{armstrong2020}. The best-fitting correlation gradients between position and velocity are listed and shown in each panel as solid lines, with dashed lines indicating the uncertainties. All three dimensions show a positive correlation between position and velocity, which is a strong indication of expansion.}
    \label{fig:vela_ob2_expansion}
\end{figure}

More recent studies have found that if associations are divided into subgroups based on their spatial and kinematic substructure then these subgroups often show evidence for expansion, even when the whole system does not. For example, \citet{kounkel2018} divided the Orion OB1 association into subgroups using spatial and kinematic information and found evidence for the expansion of the Orion D subgroup. \citet{cantat-gaudin2019} divided the young stars of the Vela OB2 association into 7 subgroups and found all of them to be expanding. \citet{armstrong2020} conducted a spectroscopic study of the centre of the Vela OB2 association to obtain radial velocities and found that, once the members of the Gamma Vel cluster were removed from the sample, the Vela OB2 association was expanding in all three dimensions (see Figure~\ref{fig:vela_ob2_expansion}).

For associations where expansion has been observed, the expansion is usually anisotropic, even when subgroups are identified using kinematics. Of the three OB associations found to be expanding by \citet{melnik2017}, two are significantly anisotropic, with only Car OB1 being consistent with isotropic expansion. \citet{wright2018} found that all three subgroups of Sco-Cen exhibit strongly anisotropic expansion, with all expanding preferentially along the Galactic $Y$ axis but not along the other two axes. \citet{cantat-gaudin2019} and \citet{armstrong2020} also both found that the expansion of the Vela OB2 association subgroups was strongly anisotropic.

In younger systems such as star forming regions or embedded clusters (which may represent the precursors of expanded associations), there is also growing evidence for expansion. \citet{kuhn2019} found evidence for expansion in $>$75\% of their sample of young clusters, including the Orion Nebula Cluster, which \citet{dario2017} had also found was expanding \citep[though][had not]{dzib2017}. \citet{kounkel2018} observed a clear radial expansion pattern in the $\lambda$ Ori cluster, while \citet{Wright2019} found strong evidence for expansion in NGC 6530, though like the older systems the expansion of NGC 6530 is highly asymmetric, with almost all the expansion occurring in the declination direction.

In summary, there is growing evidence that many, if not the majority of, associations have substructures within them that are expanding. Older studies were unable to identify this expansion due to a combination of inferior kinematic data, poorly-defined association membership lists, or because they were searching for expansion across the entire system rather than within the subgroups. The expansion that has been measured is generally anisotropic, a trend which also extends towards younger and embedded clusters.

\subsection{Ages and age spreads}
\label{sec:ages}

Establishing the ages of association members and quantifying differences, gradients and spreads in those ages are a crucial part of understanding the formation and evolution of associations. The only star with a model-independent age is the Sun, and beyond that there is a hierarchy of age-determination methods, of decreasing accuracy \citep[e.g., see][]{soderblom2010}. For young stars, these range from model-dependent methods such as asteroseismology, the fitting of stellar evolutionary isochrones in the HR, colour-magnitude or the $\log g$ vs $T_{\rm eff}$ (Kiel, or spectroscopic HR) diagrams and the depletion and diffusion of light elements, to more indirect or empirically calibrated methods such as using the time-dependence of accretion, rotation and magnetic activity \citep{soderblom2014}.

The ages of high-mass stars can be estimated from their positions in the HR or spectroscopic HR diagrams. Age discrimination is possible because O and early B-type stars evolve quickly. Accurate spectroscopic determination of the stellar parameters and a distance (for the HR diagram) are required. Typically these techniques have been deployed in more distant but rich associations, where the large number of O-stars can be used to make statistical inferences about ages and age spreads. For example \cite{wright2015} used 169 OB stars in Cyg OB2 to infer an age spread from 1--7 Myr. This work was expanded by \cite{berlanas2020}, with the benefit of {\it Gaia} DR2 parallaxes, and evidence is presented for `bursts' of star formation (or at least, bursts of O-star formation) at ages of 3 and 5 Myr. Such studies are difficult because observational errors in the stellar parameters lead to significant age uncertainties and the fidelity of these ages can be compromised by unresolved binarity and a genuine astrophysical spread in the HR diagram caused by rotation-dependent internal mixing and mass-loss rate uncertainties (for the highest mass evolved stars). The relative paucity of high-mass stars can be countered by access to the larger low-mass populations, which further allow a comparison of the ages of the high- and low-mass stars.

The low-mass populations of young ($\leq 10$ Myr) clusters and associations ubiquitously show a large spread of luminosity around any mean isochrone in the HR diagram. The causes of this spread are long-debated \citep[see][]{hillenbrand2008, jeffries2011, soderblom2014}. Observational uncertainties play a role (particularly dealing with reddening, accretion, variability and binarity), but cannot account for all or even most of the dispersion \citep[e.g.,][]{reggiani2011}. If the dispersion were attributed to age, it would equate to age spreads $\geq 10$ Myr in many clusters and associations, but it is equally possible that astrophysical scatter associated with varied accretion histories or differing levels of magnetic activity might play a significant role \citep{baraffe2017, gully2017}. This dispersion is a poorly understood nuisance factor when searching for age gradients or differences between populations within an association and hampers interpretations involving sequential, triggered or bursts of star formation \citep[though see][for an instance in the Orion Nebula cluster, where multiple star formation bursts separated by $\sim 1$ Myr may have been resolved]{jerabkova2019}.

\begin{figure*}[h]
    \centering
    \includegraphics[width = 15cm]{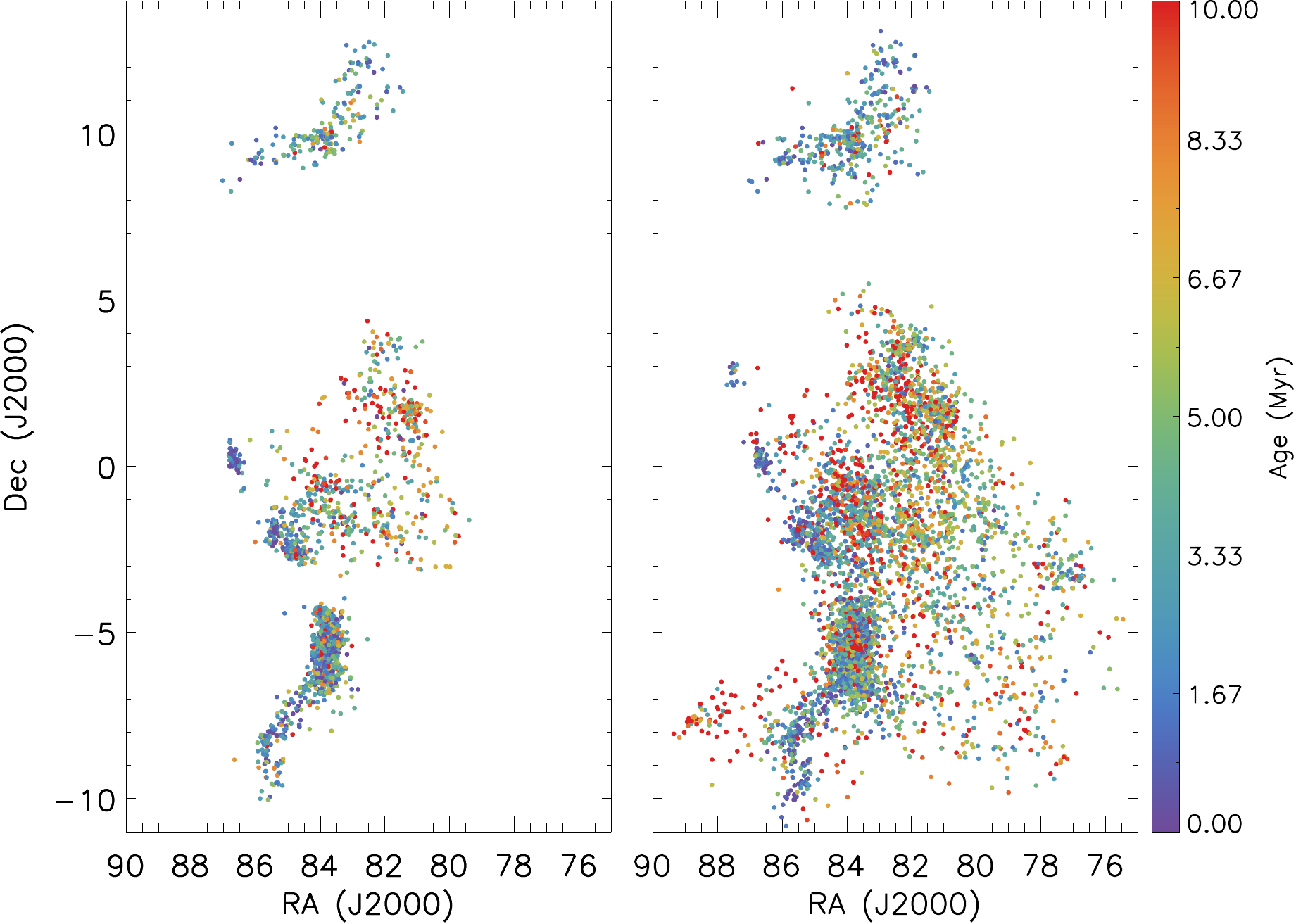}
    \caption{Estimated stellar ages across the Orion OB1 association from
\citet{kounkel2018}. Left: ages derived using spectroscopic
effective temperature and bolometric luminosities in the HR diagram. Right: ages derived using just photometry in the colour magnitude
diagram (distances assigned using the average distance to stars in each
group).}
    \label{fig:orion-age}
\end{figure*}

Despite these difficulties there has been a frenzy of activity in the past few years, exploiting both wide-field spectroscopic surveys and {\it Gaia} astrometry to probe the age structure of local associations. In Orion, \cite{da-rio2016} looked at the stellar populations along the Orion A molecular cloud, using near-IR gravity diagnostics to confirm a spread in {\it radius} and thus possibly some age spread, but found little evidence for any age gradient along the cloud. {\it Gaia} DR2 studies by \cite{kounkel2018}, \cite{zari2019} and \cite{kos2019} have dissected the wider Orion OB association using clustering algorithms in position and velocity phase-space (see Figure~\ref{fig:orion-age} for an example). The ages quoted for these sub-regions range from 1--21 Myr, spread over dimensions of $\sim$100 pc. That the temporal substructure exposed by such studies shows a correlation with the spatial and kinematic substructure in these associations provides strong evidence that this substructure is real and reflects the star formation history within the association.

A similarly complex pattern is emerging in the Sco-Cen and Vela OB2 associations. In Sco-Cen the rich, low-mass population has been used to trace clear age differences (on average) between the younger Upper Sco region at $\sim 10$ Myr and the older Upper Centaurus-Lupus and Lower Centaurus-Crux regions at $\sim 16$ Myr \citep{pecaut2016}. \cite{squicciarini2021} have used {\it Gaia} EDR3 astrometry and spectroscopic radial velocities to show that about half the Upper Sco population is in the form of a clustered population that existed in more compact configurations in the past, whilst the rest is more diffuse. They suggest star formation proceeded over about 10 Myr in small groups that gradually dissolve; indeed the deduced kinematic ages (see Section \ref{sec:kinages}) of the compact subgroups are younger than their isochronal ages. In a similar exercise for Vela OB2, \cite{cantat-gaudin2019} found seven subgroups with ages from 8 to 50 Myr. Age was not strongly correlated with position, but was correlated with kinematics suggesting a more turbulent than sequential star forming history. See Section~\ref{sec:propagation} for a more detailed discussion of the star formation history within associations.

In all these works it is explicitly (or implicitly) assumed that the model-dependent ages are accurate. Whilst there can be some confidence in the {\it relative} ages (or at least the age order) of different groups of stars, their absolute ages are more uncertain. There is a long-standing problem \citep[see][and references therein]{bell2013, pecaut2016} that more massive stars tend to have older isochronal ages than their lower mass siblings by factors of $\sim 2$. Whilst there are undoubtedly problems to solve in the high-mass stellar modelling, it seems likely, with the emergence of evidence from PMS low-mass eclipsing binary systems \citep{kraus2015, david2019} and lithium depletion \citep[][]{jeffries2017} in associations, that fitting conventional low-mass isochrones underestimates the ages (and masses) of low-mass PMS stars by factors of $1.5$--$2$ and that the adoption of models incorporating rotation, magnetic activity and starspots may bring these ages into much closer agreement with those of high-mass stars \citep[e.g.,][]{feiden2016, somers2020} -- high-mass stars can also appear younger in the HR diagram if they are born rotating significantly rapidly \citep[e.g.,][]{ekstrom2012}. These uncertainties must be considered when discussing proposed scenarios where low-mass star formation is affected or triggered by the birth or deaths of high-mass stars.

\subsection{Kinematic ages}
\label{sec:kinages}

Kinematic ages are often cited as providing a model-independent age for a group of stars, since they do not rely on any stellar physics. They are calculated from the time the system needs to expand from an initially compact configuration at its current rate to reach its present size \citep[e.g.,][]{lesh1968,blaauw1978,makarov2007}.  Their validity as a model-independent age is based on the assumption that the group of stars has expanded, unhindered, from an initially compact state to its current configuration, and that the expansion began at, or shortly after, the birth of the stars.

Kinematic ages have been calculated for many different associations and moving groups \citep[e.g.,][]{Ducourant2014}. Historically, these have often disagreed with isochronal ages calculated for those groups \citep[e.g.,][]{soderblom2010}. However, the combination of improved {\it Gaia} astrometry, a revision of the pre-MS evolutionary age scale \citep[e.g.,][]{bell2013}, and improved kinematic age calculation methods \citep[e.g.,][]{miret-roig2018,crundall2019} have resolved many discrepancies. For example, the $\beta$ Pic moving group has recently been calculated to have a kinematic age of $18.3^{+1.3}_{-1.2}$~Myr \citep{crundall2019}, in excellent agreement with its lithium depletion boundary age of $21 \pm 4$~Myrs \citep{binks2014}.

The reliability of kinematic ages is dependent on a number of assumptions \citep[see][for a general critique of this method]{soderblom2014}. Foremost amongst these is the assumption that the time at which the stars were closest to each other was when they formed. An example relevant to this is the $\lambda$ Ori system, whose isochronal age suggests it is much younger than its kinematic age \citep{kounkel2018}. If the isochronal age is correct, and not underestimated (see Section \ref{sec:ages} for a discussion on the reliability of isochronal ages), it would imply that the stars in the $\lambda$ Ori group may have formed from material that was already expanding. If this is a common occurrence it raises serious questions over the validity of kinematic ages. Related to this is the assumption that associations formed in a compact configuration (see Section \ref{sec:expansion}, and in particular the evidence for anisotropic expansion), with equally serious implications.

\citet{brown1997} performed N-body simulations to test the validity of kinematic ages and how astrometric uncertainties and the different methods employed to measure kinematic ages can affect the results. Their results suggest that the traceback method will underestimate the age of the association (and overestimate its initial size), with ages converging to $\sim$4~Myr, while the method of comparing velocity with position in a given dimension can also lead to significant uncertainties. Approaches such as the forward-modelling method used by \citet{crundall2019} have the potential to address some of these issues, but could be computationally time-consuming for large systems.

\subsection{Distribution within the Galaxy}
\label{sec:distribution}

\begin{figure*}
    \vspace{-1cm}
    \centering
    \includegraphics[width = 15cm]{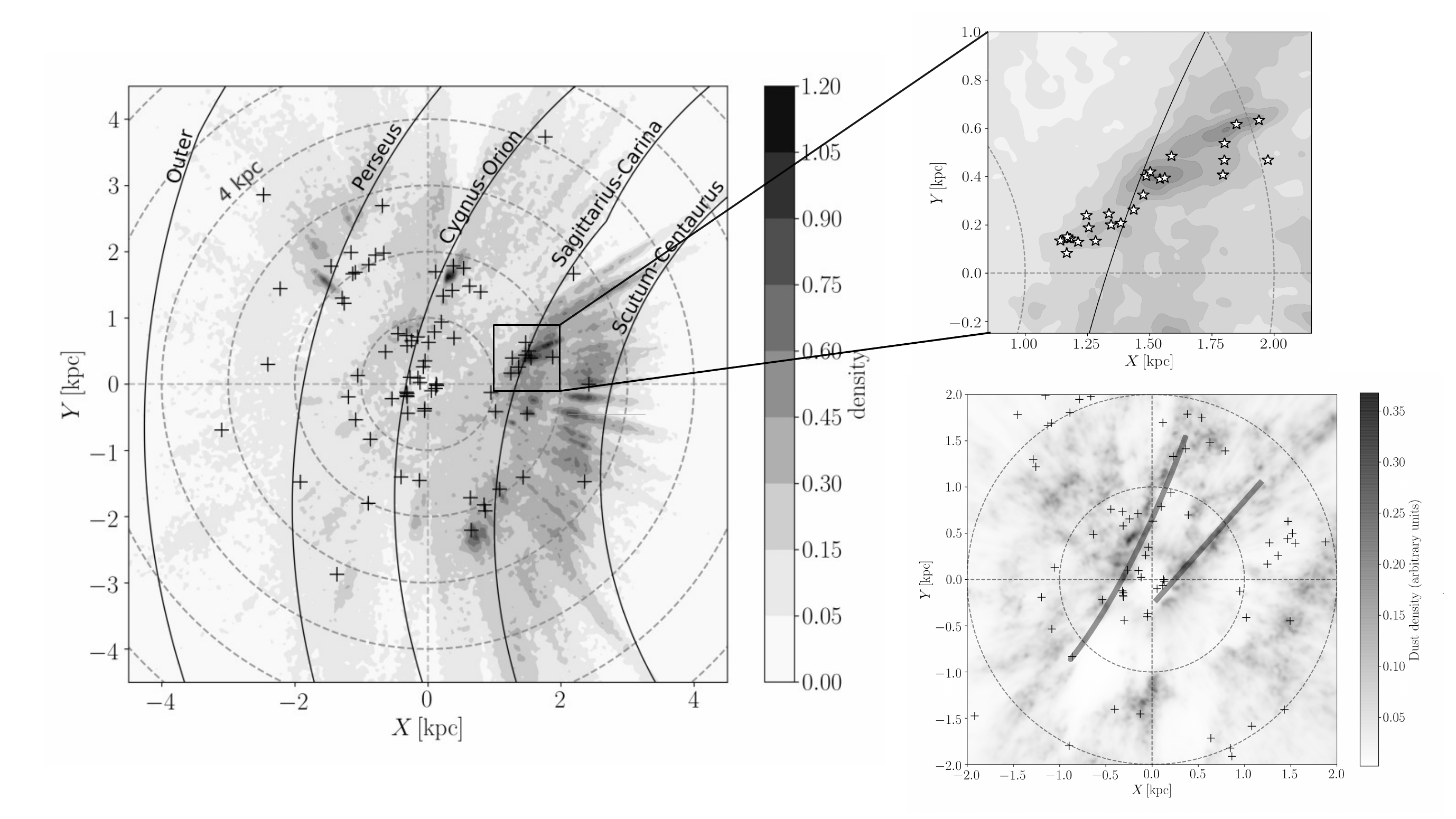}
    \caption{Left: Density distribution of OBA stars from \cite{zari2021} in the Galactic plane. The density is displayed in arbitrary units. The Sun is at $X, Y = 0,0$. The x-axis is directed towards the Galactic centre, and the y-axis in the direction of Galactic rotation. The z-axis is perpendicular to the plane. The dashed circles have radii from 1 to 6 kpc. The black crosses indicate the position of the OB associations listed in \cite{wright2020}. The black lines show the spiral arm model from \cite{reid2019}. Top Right: Zoom-in of the OBA star density map shown on the left. The white stars represent YSO groups identified by \cite{Kuhn2021a}. The black thin line shows the location of the Sagittarius-Carina arm. Bottom right: Density distribution of dust from \cite{Lallement2019}. As in the left plot, black crosses indicate the position of OB associations listed in \cite{wright2020}. The thick gray lines represent the approximate location of the Radcliffe wave (left) and the Split (right).}
    \label{fig:map_plane}
\end{figure*}

Since the seminal work by \cite{morgan1953}, OB stars and OB associations have been used to trace the spatial distribution of young stars in the Milky Way, and to probe the spiral arms of the Galaxy. In recent years many studies 
have linked the spatial distribution of young stars, young clusters, and associations in the solar neighbourhood ($d < 500-1000$~pc) into larger structures.

\cite{bouy2015} studied the spatial distribution of O and B-type stars within 500 pc of the Sun. They suggested that the distribution of OB stars in the solar neighbourhood is described by stream-like structures called `blue-streams'. Such blue-streams are associated with the three largest OB associations within 500 pc of the Sun: Orion OB1, Vela OB2 and Sco-Cen. The work of \cite{bouy2015} was based on data from the \textit{Hipparcos} satellite, and motivated \cite{zari2018} to perform a follow-up study using \textit{Gaia} DR2. \cite{zari2018} confirmed that the 3D structure of star forming regions in the solar neighbourhood is complex and filamentary, although they did not find evidence for the presence of the `blue-streams' hypothesised by \cite{bouy2015}. 
Indeed, subsequent studies have found that numerous associations in the solar neighbourhood can be linked to two gaseous structures: the Radcliffe wave and the Split. The Radcliffe wave \citep{Alves2020, Zucker2020, Green2019} is a coherent and narrow structure around 2.7 kpc long, whose 3D shape is well described by a damped sinusoidal wave in the plane of the Milky Way. The Radcliffe wave corresponds to the densest part of the Local Arm of the Milky Way and includes star forming regions in Orion, Taurus, Perseus and Cygnus. The Split \citep{Lallement2019} is argued to be a long arm segment linking the Local and Sagittarius–Carina spiral arms, and includes the Sco-Cen association. Figure \ref{fig:map_plane} (bottom right) shows a schematic representation of the Radcliffe wave and the Split, plotted on top of the dust density map from \cite{Lallement2019}.

Thanks to data from \textit{Gaia} it has been possible to characterise the distribution of associations and young stars beyond the solar neighbourhood. \cite{zari2021} estimated and studied the 3D distribution of OBA stars in the Milky Way disk within 4-5 kpc of the Sun. Figure \ref{fig:map_plane} (left) shows the distribution of the OB associations from the list presented in Table 1 of \citet[][and references therein]{wright2020} projected on the Galactic plane, plotted on top of the density distribution of the filtered sample of OBA stars from \cite{zari2021}. In general, the distribution of OB associations and OBA stars trace similar structures. These structures can be identified as part of the  Sagittarius  (or  sometimes  Sagittarius-Carina)  Arm  towards  the  inner  Galaxy,  the  Orion  (or  sometimes  Cygnus-Orion) Spur approximately at the Sun's position, and the Perseus Arm towards the outer Galaxy. There are however a number of important differences. The distribution of OBA stars shows a strong over-density corresponding to the Scutum-Centaurus arm towards the inner Galaxy (at approximately $X \sim 2$ kpc). Only two associations seem to be loosely linked with this arm suggesting our census of OB associations is incomplete at such distances. Many associations also do not seem to correspond to any significant over-densities in the OBA star distribution: such associations might have too low a density to appear on Fig. \ref{fig:map_plane} or may just be over-densities of OB stars in the sky. Finally, the mean distances to numerous associations derived in the literature do not correspond to the distances of the over-densities in Fig. \ref{fig:map_plane}, although they are towards the same lines of sight. This calls again for a revision of the census, membership, extent and distances of associations \citep[see e.g.,][]{quintana2021}.

Finally, young stellar objects (YSOs) can also be used to trace Galactic structure.
\cite{Kuhn2021b} presented a catalogue of candidate YSOs, and found groups of YSO candidates associated with the Local, Sagittarius-Carina Arm, and Scutum-Centaurus Arms. \cite{Kuhn2021a} used the same catalogue and focused on a linear feature between Galactic longitudes $l \approx 4^{\circ}–18.5^{\circ}$ including the star forming regions M8, M16, M17 and M20. Figure \ref{fig:map_plane} (top right) shows the YSO groups associated with this structure compared to the OBA star density map of \citet{zari2021} (Fig. \ref{fig:map_plane} left). The structure traced by the YSO groups corresponds to a prominent over-density that is also visible in the OBA star map and mapped out by the previously-known Sagittarius, Scutum and Serpens OB associations in this region from \citet{wright2020}. The structure does not seem to be isolated, but appears to be a feature of the Sagittarius-Carina arm, similar to those observed in external galaxies.

\section{\textbf{Discussion}}

Historically, star clusters have been much better studied than associations due to the observational bias that they are much easier to study and categorise as they are spatially much more compact, and suffer far less from back/foreground contamination. Arguably, this has led to a bias towards clusters being considered much more important than associations as sites of star formation. The frequent use of the phrase `most stars form in clusters' is usually supported by a reference to \citet{lada2003}.  However, Lada \& Lada define a `cluster' as a group of 35 or more physically related stars whose stellar mass density exceeds $1.0 M_\odot$ pc$^{-3}$  -- a broad definition which says nothing about boundedness.  A more correct statement, and one more consistent with Lada \& Lada's view, would be `most stars form at significantly higher densities than the field', followed by `and around 10\% of stars remain in bound clusters at significantly higher densities for $>10$ Myr'.

\subsection{Formation and origin}

There are many theories as to how GMCs form, and how they turn a globally supersonic medium into star-forming `units'.  For a detailed overview of star formation, and, in particular, how GMCs convert gas into stars, we direct the reader to other chapters in this volume, such as Chapters 1, 5 and 7. For the purpose of this chapter the important point is that when we observe stars with ages of a few Myr we observe them at both relatively low densities in associations and at high densities in bound clusters.  A key question in star formation is: why?

At one extreme is the view that all stars form at high densities in bound clusters, but that the vast majority of these clusters are very rapidly dispersed by gas expulsion.  The essence of this idea is that an embedded star cluster in virial equilibrium will expel the residual gas left over from star formation thanks to feedback from massive stars on a timescale of a few Myr. Since the star formation efficiency is generally quite low and therefore the majority of the mass is still in the form of gas, expelling this gas will dramatically reduce the potential of the cluster causing it to unbind and quickly disperse \citep[e.g.][]{hills1980,kroupa2001,goodwin2006}.  Associations are then the expanding remnants of one (or more) originally dense star clusters\footnote{For much of the 1990s and 2000s this was one of the dominant pictures of star formation and an underlying assumption in much work on clusters e.g. \citet{goodwin2006}.}.

It has become clear that the view of associations as the remains of (a small number of) expanded clusters does not fit the observational data. Recent studies have revealed that associations are more spatially extended than previously thought, and that they present a high degree of substructure in physical space, kinematics, and age (see Section 3). This suggests that associations are globally {\em dynamically} very young\footnote{Note that dynamical age and chronological age can be very different. The former is a measure of age in terms of crossing times, which, in unbound systems, is effectively infinite (some relaxation can occur on small scales and within substructures, and unbound initially high density regions will have encounters early-on that can erase substructures, but the point stands).} \citep{parker2014}, since any form of substructure is easier to erase than to form -- any encounters within a group of stars will erase substructure within a few crossing times, which can be particularly rapid in bound stellar systems \citep[e.g.,][]{goodwin1997,goodwin2004,parker2014}. This strongly suggests that associations form with a similar spatial configuration as we currently observe them: over large volumes, with considerable substructure, at low average densities (though some stars may form at higher densities in subclusters), and most likely globally unbound. 

The kinematic and physical substructure observed in associations reflects that of GMCs. Observations of cores within GMCs show stars often form at relatively low densities with significant substructure.  The detailed large-scale sub-mm maps from {\em Herschel} show cores (unsurprisingly) following the complex gas structures.  Two particularly good examples are Aquila \citep[][see their fig. 1]{konyves2010}, and Orion B \citep[][see their fig. 5]{konyves2020}.  Similar structures are seen in very young class I/II stars traced by e.g. {\em Spitzer} \citep{gutermuth2009}. We would expect the distributions of cores and stars we see over scales of several pc in e.g. Aquila and Orion B to be only marginally bound or unbound \citep[e.g. from Larson's laws,][]{larson1981}.  Therefore, it is difficult to imagine these regions evolving into anything other than associations (with maybe some small subclusters).

\subsection{Propagation of star formation in OB associations}
\label{sec:propagation}

Whilst complex and difficult to measure, the ages of stars and age distributions within associations provide clues as to how star formation has proceeded in different regions.  

The classical model for the formation of OB sub-groups in associations was proposed by \cite{elmegreen1977}. This model predicts that ionising radiation and winds from massive stars terminate the star formation process and drive shocks in other parts of the parental cloud. New generations of massive stars are born, and the process is repeated. As a consequence, low-mass stars should be older than OB stars, since only OB stars are formed by triggering, while low-mass stars should form spontaneously through the cloud. Another popular model is based on the mechanism of radiation driven implosion \citep{Sandford1982, kessel2003, bisbas2011, haworth2012}. According to this model, low-mass stars should be younger than the OB stars (which initiate  their formation), and one may expect to see an age gradient in the low-mass population \citep{preibisch2007}. However, from an observational perspective, there is currently no convincing evidence for either of these mechanisms.

Being the closest OB association to the Sun ($d\sim140$~pc), Sco-Cen has provided an ideal laboratory for testing theories of the propagation of star formation \citep{preibisch1998, preibisch2002, preibisch2007}. Following the pre-\textit{Gaia} study by \citet{pecaut2016} and \textit{Gaia} studies mentioned above, \cite{krause2018} and \cite{kerr2021} proposed more complex models for the star formation history of Sco-Cen, which take into account the increasing layers of complexity emerging from the data. \cite{krause2018} combined gas observations and hydrodynamical simulations to study the formation of the Sco-Cen super bubble, and suggested the following scenario for the evolution of the association. Dense gas was originally distributed in  an  elongated  cloud,  which  occupied  the  current area of the association. The star-formation events in Upper Centaurus Lupus and Lower Centaurus Crux lead to  super-bubbles that expanded, surrounding and compressing the parental molecular cloud, and triggering star formation in  Upper Scorpius.  This  scenario  predicts  the  formation  of  kinematically-coherent  sub-groups  within  the  association that  move in different directions, similar to the observed kinematics in Sco-Cen \citep{wright2018}.  \cite{krause2018} also predict that young groups could also be found in regions containing older stars,  and  that  several  young  groups  with similar ages might form over large scales. This is consistent with what is observed in Sco-Cen and other associations. The ages derived by \cite{kerr2021} however seem inconsistent with the model proposed by \cite{krause2018}. \cite{kerr2021} argue that Krause's `surround and squash' scenario might act in certain regions of the association, while in others a simple model of sequential star formation might have been enough for the propagation of star formation. It could therefore be that multiple methods of star formation propagation have acted within a single association.

The shell-like distribution of stars and gas in the Vela OB2 complex suggests an episode of triggered star formation that shaped the gas into a shell, causing it to compress and form stars \citep{cantat-gaudin2019b}. This would indicate that the expansion of the IRAS Vela shell preceded the formation of the stars in Vela OB2, with the expanding motion of the gas imprinted on the stars that formed. The mechanism responsible for shaping the distribution of stars and gas would have to be energetic enough to not only compress gas into forming stars, but also induce the expansion of the IRAS Vela shell, which \citet{cantat-gaudin2019b} suggest was most likely a supernova.

A similar conclusion was drawn in Orion by \citet{kounkel2020} and \citet{grossschedl2021}, who independently concluded that a major feedback event occurred in the region $\approx$~6~Myr ago. Such an event shaped the spatial distribution and kinematics of the gas and young stars that are observed today, and possibly triggered the formation of some of the youngest stellar groups in the region. Although \citet{kounkel2020} and \citet{grossschedl2021} studied the motions of different groups as a function of time, they did not assess the presence of age gradients at the time such groups were formed (instead assessing them at the current time). This may be important to better explain the formation history of associations and potentially to evaluate the importance of triggering. The star formation history of the Orion OB association is however not completely understood. The age distribution of the groups identified for instance by \cite{kounkel2018} and \cite{zari2019} has not produced a clear picture of the progression of star formation across the association.

It is interesting to compare the large age spreads and seemingly complex star formation histories of associations with those of clusters.  In particular, the association Cyg OB2 and the young massive cluster Wd 1 are both a few $\times 10^4 M_\odot$ with an age of $\sim 5$ Myr \citep{brandner2008,wright2015}.  However, Cyg OB2 seems to have a significant age spread \citep{wright2015}, whilst Wd 1 seems to have formed in a single burst of star formation \citep{brandner2008}.  Presumably the very different star formation histories are telling us about the properties of the (presumably fairly similar mass) GMCs from which they formed. The reason why some GMCs produce massive OB associations, while other GMCs of similar mass produce massive clusters is, however, still far from clear.

\subsection{Dynamical evolution and dispersal}

OB associations are massive, extended stellar structures whose low density suggests they are globally unbound.  Their expansion rates are sufficiently low (or are limited to one or two dimensions) that they remain co-moving for some length of time and form elongated structures. Many elongated, and relatively old, association-like structures have been discovered in recent years using {\it Gaia} data.

\cite{jerabkova2019} reported the  discovery of a $\approx$~17~Myr population, clustered in proper motions, but filamentary and extending for $\approx$90~pc in physical space, which they interpret as a relic of star formation in a molecular cloud filament. An analogous structure was also reported in Vela by \cite{beccari2020}, extending $\sim$260~pc and with an approximate age of 35~Myr. \citet{kounkel2019} also identify a number of filamentary structures orientated parallel to the Galactic Plane and with ages up to $\sim$100-200 Myr (young enough that they have not completed a full Galactic orbit) -- co-moving, filamentary and extended groups much older than this appear difficult to identify using currently-available data and methods \citep{kounkel2019}.

These groups are too young to have been significantly affected by tidal forces and therefore their extended structure is likely to be mostly primordial. As associations are thought to form from filamentary molecular clouds, the resulting young population should retain a similar string-like morphology, which would explain the observation of these extended structures. Furthermore, since the dissolution of associations is not completely isotropic (see Section~\ref{sec:expansion}), and in some cases shows preferential expansion in the X and Y directions of the Galactic Plane \citep[e.g.,][]{wright2018}, this may explain the origin of such extended and filamentary structures.

As a population of stars is orbiting around the Galaxy, its velocity dispersion increases due to dynamical evolution, tidal forces (such as from passing stars, nearby GMCs, and the Galaxy itself) and shear due to Galactic rotation. Over time stars in the group will slowly drift away and eventually most stars in the group will have dispersed into the galactic field population. Some of the stars may still remain as a coherent group for several tens if not hundreds of Myr, with the populations that were originally more massive managing to retain a larger fraction of their members as co-moving for longer periods of time. \citet{kounkel2019} identified a large number of co-moving groups in the Milky Way using {\it Gaia} DR2 data. They found that the number of stars within a co-moving and coherent group decreases with time according to a power law.

\section{\textbf{Implications}}

Over the last decade there have been many changes to our view of OB associations and their origins, and these have significant implications for our understanding of the star and planet formation processes.

\subsection{At what densities does star formation take place?}

A key question for star formation studies in galaxies is in what sort of environments do stars typically form?  Do most stars form at the fairly low stellar densities observed in associations?  Or do they often form at higher (cluster-like) densities and disperse?  Or are both important \citep[e.g.][]{bressert2010,kruijssen2012}? This is a question with potentially significant implications for star formation: do stellar systems form in cores that are essentially `isolated', or do dynamics and encounters play an important role in the formation of stars and mass assembly?

The distinction is typically made as to whether a particular region is a `cluster' or an `association' from its total energy: clusters are bound, and associations are unbound \citep[e.g.][]{gieles2011}. However, from the point of view of star formation, multiple systems or planets, a more useful distinction is if stellar systems have ever `known' about other systems. This relies on their `density history' -- in particular, the highest density environment they have ever been in, and for how long.

There is a `critical' density in a star forming environment of $\sim$100 $M_\odot$ pc$^{-3}$ above which star forming cores interact with each other.  This comes roughly from how closely one can pack 0.1 pc radius cores with a lifetime of $\sim $0.2 Myr and velocity dispersion of $\sim$1 km s$^{-1}$ \citep{goodwin2007} -- above this density cores will interact with each-other while forming stars and not be `isolated'.  Even if a region at this density is unbound, encounters will occur before it expands (unless the expansion velocity is 10s km s$^{-1}$).

If, as seems clear from the data on associations, most of the stars in associations have always been at low density, then these systems will be `pristine'.  That is, they have not been altered by external encounters/irradiation (however, secular effects may be important) and it will be important to compare the properties of such stars, together with their circumstellar environments and planetary systems with those found in dense clusters.

\subsection{Do binary and multiple systems evolve differently in clusters and associations?}

A significant fraction of stars are thought to form in binary and higher-order multiple systems \citep[see reviews by][]{goodwin2007,reipurth2014,duchene2013}.  Binary and multiple systems can be changed or destroyed by secular decay (for higher-order multiple systems) and by external encounters. Secular decay should be environmentally-independent (although small external perturbations could play some role) and occurs on order of the multiple system's crossing time (typically 10s to 100s kyr). For example, a triple system would usually be expected to decay to a (hardened) binary and a single star \citep[e.g.][]{goodwin2005,reipurth2010}. The effect of close encounters depends on both the frequency and energy of encounters \citep{heggie1975,hills1975,kroupa1995} -- both of which are density-dependent \citep[though it is generally a very stochastic process,][]{parker2012b}. The more encounters and the more energetic they are, the more systems we would expect to be destroyed (or at least significantly altered).

The effect of both secular decay and close encounters is therefore to decrease the multiplicity fraction (MF) of a group of stars over time. Unfortunately, these two effects are extremely difficult to disentangle as they both lead, in a very stocastic way, to a reduction in the MF.

Observations seem to show that multiplicity decreases with age. The MF for Class 0 YSOs is close to unity \citep{chen2013,tobin2019}, but by the time these sources reach the Class I and Class II stage the MF is closer to what is typically observed in the field \citep[see][]{duchene2013,reipurth2014}. However, the simplifying assumption of a single initial population of multiple systems which are then altered by dynamics \citep[e.g.][]{kroupa1995,kroupa2001} seems difficult to justify {\it a priori}.  This is because if the star formation environments are different enough to produce an association versus a cluster, one might well expect that the initial multiplicity properties of stars would be different.  This leaves us with attempting to reverse-engineer from current populations the stochastic history of secular decays and external encounters on potentially different initial multiple populations which could give rise to the currently observed populations.

From our earlier discussions that most stars in associations seem always to have been at low density, while stars in clusters we know have spent a significant fraction (if not all) of their lives at high density, we would expect the multiplicity properties of clusters and associations to be different as the encounter/irradiation histories of stars within them were different. There is some evidence supporting this view. In the low-density Taurus Molecular clouds the multiplicity fraction is very high, $\sim$66--75\%, in excess of what is found in the field by a factor of $\sim$2 for stars of a comparable mass. In contrast, in the dense Orion Nebula Cluster (ONC), the multiplicity fraction at separations of 67.5--675 AU is consistent with the field \citep{reipurth2007}.

The dependence of the MF on both local stellar density and binary/multiple system separation is also interesting, with wide ($>$1000 AU) binaries particularly sensitive to environment. For example in the low-density Taurus Molecular cloud and Cygnus OB2 association there are many such wide binaries \citep{kraus2011,caballero-nieves2020}, while in the dense ONC no binaries with separations over 1000~AU have been conclusively identified \citep{scally1999}. These observations can help place limits on the density history of such binaries. For example, \citet{griffiths2018} showed that the high fraction of wide binaries amongst massive stars in Cyg OB2 was inconsistent with them being born in a dense cluster with other massive stars, because wide binaries in clusters will be destroyed or hardened by the presence of another massive star.

For close binaries ($<$100 AU), all nearby low-density regions that have been studied appear to show a MF that is very similar to the field for separations of 60--100 AU (with only a tentative hint of a density dependence) but an excess of systems $< 60$ AU by a factor of roughly two \citep{king2012}. Separations of $< 60$ AU are interesting as these systems are `hard' in all local environments and so should survive into the field. So an excess in local regions over the field of a factor of two suggests some environments must under-produce fairly close systems.  The obvious suspect for under-production are clusters. The ONC is the only dense cluster close enough for detailed multiplicity studies, and it has been difficult to resolve close ($<60$ AU) companions in the ONC.  A recent search for close companions in the ONC using high resolution imaging has shown a similar excess in the MF at 20--60 AU as in low density regions: twice the field \citep{duchene2018}. However, this effect may be limited to particular primary mass ranges \citep{de-furio2019,de-furio2021}. It is also problematic to extrapolate from a sample of one cluster.

At intermediate separations (100-1000 AU) the picture is still unresolved. There is evidence that stars in the denser regions of the Orion Molecular Clouds have a higher MF at separations of 100--1000 AU than the regions with lower stellar density \citep{kounkel2016,tobin2022}. Thus, despite being presumably much more sensitive to destruction via encounters than low density populations, clustered regions can seemingly produce a larger fraction of companions at some separation ranges. This could be primordial, it could be due to the tightening of previously wider binaries, or it could be due to the loose capture of stars into multiple systems \citet{moeckel2011}.

In summary it is currently impossible to say with any certainty if or how birth multiplicity varies with environment (or if it is universal) and how it is processed differently in different environments). In the future, in light of the larger census of YSOs that can now be assembled with a variety of different surveys such as {\em Gaia}, and with new facilities coming online that offer increasingly better spatial resolution, it may be possible to further improve multiplicity statistics in a larger number of star forming regions to better determine the role of the environment in the underlying processes governing both the initial multiplicity of a population and its secular decay and the effect of external encounters.

\subsection{Do protoplanetary disks and planetary systems evolve differently in clusters and associations?}

There are two external effects that can alter the properties of planetary systems: encounters and irradiation. Encounters can affect both disks and ($N$-body) planetary systems, while irradiation only affects disks. The stellar density of the environment surrounding a young star (and the evolution of that density) determines both of these effects through the likelihood of close encounters and the local radiation and particle field.

In dense clusters, disks may be photoevaporated by EUV radiation from nearby O-type stars \citep[typically when within 1~pc, e.g.,][]{winter2018b,cai2019,parker2021}, while even a brief period ($< 1$ Myr) at high density is enough to alter planetary system properties \citep[see][and references therein]{parker2020}. At lower densities in associations these effects should be different and reduced in comparison \citep[e.g.][]{laughlin1998,clarke1993,adams2006,wright2012,portegies-zwart2015, vincke2015, nicholson2019,winter2019}.

Current observational evidence is limited, but suggests that disks are affected by their environment. \citet{dejuanoverlar2012} found evidence for a decrease in the protoplanetary disk radius at very high cluster densities, but their sample of sources at such high densities was small. \citet{eisner2018} made a similar discovery in the ONC, when compared to lower-density star forming regions. \citet{guarcello2016} found that the fraction of stars with protoplanetary disks is inversely proportional to the local ultraviolet flux in Cyg~OB2, suggesting disk dissipation is dominated by photo-evaporation and therefore strongly dependent on environment.

The extreme mass loss rate in the protoplanetary disks in the ONC from photoevaporation from $\theta^1$ Ori C \citep{henney2002} has raised the question of how these disks can survive for a prolonged period of time to be currently detected, without invoking implausibly large initial disk masses or unrealistic ages for $\theta^1$ Ori C, given the typical ages of other stars in the vicinity. However, the age gradient of stars in the ONC \citep{beccari2017} may provide a solution to the problem: older stars have had a greater shielding from photoevaporation through the surrounding dust, allowing them to circulate throughout the cluster and migrate away from $\theta^1$ Ori C. On the other hand, the most strongly irradiated stars are the youngest, with the greatest mass reservoir remaining \citep{winter2019a}.

Further measurements of protoplanetary disk frequencies in OB associations will be needed to fully understand the influence of environment on disk evolution. To understand how the environment affects the properties (or frequency of) planetary systems will require large samples of planetary systems in star clusters and OB associations to be compiled. The current evidence and our understand of planet formation does suggest that the properties of planetary systems should depend on environment, though the exact form of this dependency could take many forms.

\section{\textbf{Future Prospects}}

Many upcoming facilities and instruments have the potential to significantly advance our understanding of associations. Here we discuss these observatories, as well as the prospects of future {\it Gaia} data releases.

\subsection{ Future {\it Gaia} data releases}

The data from the \textit{Gaia} satellite have allowed for significant advancements in our understanding of associations. Thanks to \textit{Gaia}'s exquisite astrometric precision, it has been possible to identify low-mass members of nearby associations (within 500~pc) and to study in detail their distribution in 3D physical space, their kinematics and ages.

With \textit{Gaia} EDR3 the precision of parallaxes and proper motions has improved on average by 20\% and by a factor of 2 respectively, compared to DR2. \textit{Gaia} DR3 (expected mid-2022) will further provide new and improved radial velocities (for $\sim$30 million stars out to $G\sim15-15.5$~mag, with precisions $<$1~km~s$^{-1}$ for the brightest stars) and astrophysical parameters for sources based on the BP/RP prism and RVS spectra (for the full content of \textit{Gaia} DR3 see \url{https://www.cosmos.esa.int/web/gaia/release}). {\it Gaia} DR4 will feature parallaxes that are more precise by a factor 1.7 with respect to \textit{Gaia} DR2, while for proper motions the gain will be a factor of 5.2 \citep{brown2021}.

The parallax precision improvement from future {\it Gaia} data releases, combined with a better understanding and treatment of the systematics, should allow the identification of pre-main sequence members of associations beyond 500 pc from the Sun, and potentially out to 2~kpc (for the most massive and brightest PMS stars). A major improvement in the study of nearby associations will come from the radial velocities provided in \textit{Gaia} DR3. This will be a uniform data set that will facilitate studies of the kinematic and dynamic properties of both the clustered and diffuse populations of associations. For example, between 40--50\% of the $\approx$ 40,000 pre-main-sequence star candidates published by \citet{zari2018} will have radial velocities in {\it Gaia} DR3, compared to the current value of 2\%.

\subsection{Multi-object spectroscopic surveys}

Spectroscopy is important in the study of associations for verifying the youth of candidate young stars (Section~\ref{sec:confirming}), providing accurate stellar parameters (allowing stars to be placed in the HR diagram and stellar ages estimated), measuring precise radial velocities (to facilitate full 3D kinematic studies) and abundances (for chemical tagging).

The next-generation of multi-object spectroscopic instruments will be ideal for studying associations due to their combination of wide fields of view (typically several square degrees) and large numbers of fibres. This will allow them to observe hundreds or thousands of stars at a time spread over large areas, thus making them ideally suited to the study of associations.

The first such facility to come online is SDSS-V \citep{Kollmeier2017}, whose northern and southern hemisphere telescopes have already embarked on am ambitious survey of the Milky Way and the Local Volume, including many thousands of both high- and low-mass young stars. The telescopes operate in both low- ($R = 2000$) and high-resolution ($R = 20,000$) in the optical and near-infrared, with $\sim$500 fibres distributed over 2.8 (northern) and 7 (southern) square degree fields of view.

WEAVE \citep[WHT Enhanced Area Velocity Explorer,][]{dalton2018} is the new multi-object spectrograph being installed on the William Herschel Telescope, with first light expected late 2022. WEAVE has a 3 square degree field of view and $\sim$1000 fibres that can observe in low- ($R = 5000$) or high-resolution ($R = 20,000$) modes. WEAVE will undertake 8 pre-planned surveys during its first 5 years of operation, including surveys targeting dispersed young stars and associations in the Galactic Plane, and open clusters.

In 2023, these instruments will be joined by ESO's 4MOST \citep[4-metre Multi Object Spectroscopic Telescope,][]{dejong2019} spectrograph, which has a 4 square degree field of view and $\sim$2400 fibres that observe in both low- ($R = 5000$--$7500$) and high-resolution ($R = 20,000$) modes synchronously. 4MOST will also operate surveys on a 5 year timescale, with multiple surveys of young stars, clusters and associations in the first slew of surveys.

\subsection{X-ray telescopes}
\label{sec:futurex-ray}

The eROSITA X-ray telescope was launched in 2019 and is performing a survey of the whole sky over 4 years, in the form of 8 successive all-sky maps taken over 6 months. The accumulated sensitivity to 0.2-8 keV X-rays will be about 25 times that of previous all-sky X-ray surveys \citep{predehl2021}. This has the potential to identify many candidate young stars across the broad areas covered by associations, by virtue of their magnetic and accretion-related activity (see Section~\ref{sec:identifying}).

The first results from eROSITA, based on the first scan of the sky, have now been released and a first look at the Sco Cen OB association was presented by \cite{Schmitt2021}. Already the sensitivity is good enough to detect M-type association members at $\sim 120$ pc with saturated levels of X-ray emission. Extrapolating to end-of-survey sensitivities, one can estimate that most low-mass PMS stars ($0.2 \leq M/M_\odot \leq 2$) will be identified out to $\sim 500$ pc and solar-type PMS stars should be detected well beyond 1 kpc. Since X-ray selection is unbiased by kinematic considerations, it is likely that these data will play a very important role in the study of the demography and structure of nearby associations over the coming decades.

 \section{Summary}
 
OB associations are important objects to study as they represent the transition phase between clustered (or unclustered) star formation and the population of mature stars orbiting in the Milky Way. As such they are important not just for understanding the evolution of stellar clustering, but also provide samples for studies of stellar evolution (at both low and high masses) and the evolution of multiple systems, protoplanetary disks, and young planetary systems.
 
Identifying members of associations in a reliable and unbiased fashion is difficult due to their low densities and the large areas of sky they cover. Candidate young stars are often identified using a combination of optical and infrared photometry, astrometry, X-ray observations and/or variability. Validation of youth can be achieved using spectroscopy, by confirming the effective temperatures of high-mass stars or by measuring either the abundance of atmospheric lithium or surface-gravity sensitive spectral features in low-mass stars. Once a reliable sample of young stars has been obtained, they can be divided into groups or sub-groups using multi-dimensional (spatial and kinematic) data and a variety of clustering algorithms.

Recent work has uncovered significant levels of spatial, kinematic and temporal substructure in associations. This substructure has helped resolve the large-scale structure of associations, explain kinematic anisotropy and complexity, and resolve age spreads commonly observed in associations. Evidence for the expansion of associations has been mixed, with early studies finding no or limited evidence for their expansion, while more recent studies (that break down the structure of associations using spatial and kinematic information) finding that expansion is a common feature of associations, though this expansion can be quite anistropic. This in particular has implications for whether kinematic ages can be used as a model-free age indicator. The large-scale star formation history within (and between) associations does not show any obvious patterns that would suggest a single mechanism responsible for the propagation of star formation (e.g., triggering of some sort) and it may be that multiple mechanisms are at work at the same time. As associations disperse they seem to form extended, filamentary structures that can survive as coherent moving groups for up to $\sim$300~Myr. The origin of their filamentary morphologies may be a combination of primordial structure and the observed anisotropic expansion patterns.

These observations suggest a picture in which associations form as extended, highly substructured and with a low average density (albeit with some over-densities). Some of the over-densities may be dense enough to form clusters, and some of these may survive as long-lived open clusters, while other expand and disperse, often anisotropically, thereby preserving an extended and filamentary structure for several hundreds of Myr.

The low densities at which many stars in associations form has implications for the star formation process and the formation and evolution of binary and multiple systems, protoplanetary disks, and young planetary systems. Full observational tests of the predictions made by such implications have yet to be carried out, but preliminary studies suggest some differences in the products of star formation between stars in dense clusters and low-density associations.

\bigskip
\textbf{Acknowledgments.} NJW acknowledges an STFC Ernest Rutherford Fellowship (grant number ST/M005569/1) and a Leverhulme Trust Research Project Grant (RPG-2019-379).

\bigskip
{\small
\bibliographystyle{pp7.bst}
\bibliography{refs.bib}

\begin{thebibliography}{228}
\parskip=0pt \itemsep=0pt \small \baselineskip=11pt
\providecommand{\natexlab}[1]{#1}

\bibitem[\protect\astroncite{\emph{{Adams} et~al.}}{2006}]{adams2006}
{Adams} F.~C. et~al. (2006) \emph{\apj}, \emph{641}, 1, 504.

\bibitem[\protect\astroncite{\emph{{Alves} et~al.}}{2020}]{Alves2020}
{Alves} J. et~al. (2020) \emph{\nat}, \emph{578}, 7794, 237.

\bibitem[\protect\astroncite{\emph{{Ambartsumian}}}{1947}]{ambartsumian1947}
{Ambartsumian} V.~A. (1947) \emph{{The evolution of stars and astrophysics}},
  Armenian Acad. of Sci.

\bibitem[\protect\astroncite{\emph{{Ambartsumian}}}{1949}]{ambartsumian1949}
{Ambartsumian} V.~A. (1949) \emph{\azh}, \emph{26}, 3.

\bibitem[\protect\astroncite{\emph{{Armstrong} et~al.}}{2018}]{armstrong2018}
{Armstrong} J.~J. et~al. (2018) \emph{\mnras}, \emph{480}, 1, L121.

\bibitem[\protect\astroncite{\emph{{Armstrong} et~al.}}{2020}]{armstrong2020}
{Armstrong} J.~J. et~al. (2020) \emph{\mnras}, \emph{494}, 4, 4794.

\bibitem[\protect\astroncite{\emph{{Arnold} et~al.}}{2020}]{arnold2020}
{Arnold} B. et~al. (2020) \emph{\mnras}, \emph{495}, 3, 3474.

\bibitem[\protect\astroncite{\emph{{Baraffe} et~al.}}{1998}]{baraffe1998}
{Baraffe} I. et~al. (1998) \emph{\aap}, \emph{337}, 403.

\bibitem[\protect\astroncite{\emph{{Baraffe} et~al.}}{2017}]{baraffe2017}
{Baraffe} I. et~al. (2017) \emph{\aap}, \emph{597}, A19.

\bibitem[\protect\astroncite{\emph{{Beccari} et~al.}}{2017}]{beccari2017}
{Beccari} G. et~al. (2017) \emph{\aap}, \emph{604}, A22.

\bibitem[\protect\astroncite{\emph{{Beccari} et~al.}}{2020}]{beccari2020}
{Beccari} G. et~al. (2020) \emph{\mnras}, \emph{491}, 2, 2205.

\bibitem[\protect\astroncite{\emph{{Bell} et~al.}}{2013}]{bell2013}
{Bell} C. P.~M. et~al. (2013) \emph{\mnras}, \emph{434}, 1, 806.

\bibitem[\protect\astroncite{\emph{{Berlanas} et~al.}}{2019}]{berlanas2019}
{Berlanas} S.~R. et~al. (2019) \emph{\mnras}, \emph{484}, 2, 1838.

\bibitem[\protect\astroncite{\emph{{Berlanas} et~al.}}{2020}]{berlanas2020}
{Berlanas} S.~R. et~al. (2020) \emph{\aap}, \emph{642}, A168.

\bibitem[\protect\astroncite{\emph{{Bica} et~al.}}{2008}]{bica2008}
{Bica} E. et~al. (2008) \emph{\mnras}, \emph{389}, 2, 678.

\bibitem[\protect\astroncite{\emph{{Binks} and {Jeffries}}}{2014}]{binks2014}
{Binks} A.~S. and {Jeffries} R.~D. (2014) \emph{\mnras}, \emph{438}, 1, L11.

\bibitem[\protect\astroncite{\emph{{Bisbas} et~al.}}{2011}]{bisbas2011}
{Bisbas} T.~G. et~al. (2011) \emph{\apj}, \emph{736}, 2, 142.

\bibitem[\protect\astroncite{\emph{{Blaauw}}}{1946}]{blaauw1946}
{Blaauw} A. (1946) \emph{Publications of the Kapteyn Astronomical Laboratory
  Groningen}, \emph{52}, 1.

\bibitem[\protect\astroncite{\emph{{Blaauw}}}{1952}]{blaauw1952}
{Blaauw} A. (1952) \emph{\bain}, \emph{11}, 414.

\bibitem[\protect\astroncite{\emph{{Blaauw}}}{1964}]{blaauw1964}
{Blaauw} A. (1964) \emph{\araa}, \emph{2}, 213.

\bibitem[\protect\astroncite{\emph{{Blaauw}}}{1978}]{blaauw1978}
{Blaauw} A. (1978) \emph{{Internal Motions and Age of the Sub-Association Upper
  Scorpio}}, p. 101, Armenian Acad. of Sci.

\bibitem[\protect\astroncite{\emph{{Blaauw}}}{1991}]{blaauw1991}
{Blaauw} A. (1991) in: \emph{The Physics of Star Formation and Early Stellar
  Evolution}, vol. 342 of \emph{NATO Advanced Study Institute (ASI) Series C},
  (edited by C.~J. {Lada} and N.~D. {Kylafis}), p. 125.

\bibitem[\protect\astroncite{\emph{{Blaha} and {Humphreys}}}{1989}]{blaha1989}
{Blaha} C. and {Humphreys} R.~M. (1989) \emph{\aj}, \emph{98}, 1598.

\bibitem[\protect\astroncite{\emph{{Bouy} and {Alves}}}{2015}]{bouy2015}
{Bouy} H. and {Alves} J. (2015) \emph{\aap}, \emph{584}, A26.

\bibitem[\protect\astroncite{\emph{{Bouy} et~al.}}{2014}]{bouy2014}
{Bouy} H. et~al. (2014) \emph{\aap}, \emph{564}, A29.

\bibitem[\protect\astroncite{\emph{{Brandner} et~al.}}{2008}]{brandner2008}
{Brandner} W. et~al. (2008) \emph{\aap}, \emph{478}, 1, 137.

\bibitem[\protect\astroncite{\emph{{Bressert} et~al.}}{2010}]{bressert2010}
{Bressert} E. et~al. (2010) \emph{\mnras}, \emph{409}, 1, L54.

\bibitem[\protect\astroncite{\emph{{Brice{\~n}o} et~al.}}{2005}]{Briceno2005}
{Brice{\~n}o} C. et~al. (2005) \emph{\aj}, \emph{129}, 2, 907.

\bibitem[\protect\astroncite{\emph{{Brice{\~n}o} et~al.}}{2019}]{Briceno2019}
{Brice{\~n}o} C. et~al. (2019) \emph{\aj}, \emph{157}, 2, 85.

\bibitem[\protect\astroncite{\emph{{Brown}}}{2021}]{brown2021}
{Brown} A. G.~A. (2021) \emph{\araa}, \emph{59}.

\bibitem[\protect\astroncite{\emph{{Brown} et~al.}}{1997}]{brown1997}
{Brown} A.~G.~A. et~al. (1997) \emph{\mnras}, \emph{285}, 3, 479.

\bibitem[\protect\astroncite{\emph{{Brown} et~al.}}{1999}]{brown1999}
{Brown} A.~G.~A. et~al. (1999) in: \emph{The Origin of Stars and Planetary
  Systems}, vol. 540 of \emph{NATO Advanced Study Institute (ASI) Series C},
  (edited by C.~J. {Lada} and N.~D. {Kylafis}), p. 411.

\bibitem[\protect\astroncite{\emph{{Buckner} et~al.}}{2019}]{buckner2019}
{Buckner} A. S.~M. et~al. (2019) \emph{\aap}, \emph{622}, A184.

\bibitem[\protect\astroncite{\emph{{Caballero-Nieves}
  et~al.}}{2020}]{caballero-nieves2020}
{Caballero-Nieves} S.~M. et~al. (2020) \emph{\aj}, \emph{160}, 3, 115.

\bibitem[\protect\astroncite{\emph{{Cai} et~al.}}{2019}]{cai2019}
{Cai} M.~X. et~al. (2019) \emph{\mnras}, \emph{489}, 3, 4311.

\bibitem[\protect\astroncite{\emph{{Cantat-Gaudin}
  et~al.}}{2019{\natexlab{a}}}]{cantat-gaudin2019b}
{Cantat-Gaudin} T. et~al. (2019{\natexlab{a}}) \emph{\aap}, \emph{621}, A115.

\bibitem[\protect\astroncite{\emph{{Cantat-Gaudin}
  et~al.}}{2019{\natexlab{b}}}]{cantat-gaudin2019}
{Cantat-Gaudin} T. et~al. (2019{\natexlab{b}}) \emph{\aap}, \emph{626}, A17.

\bibitem[\protect\astroncite{\emph{{Carpenter}}}{2000}]{carpenter2000}
{Carpenter} J.~M. (2000) \emph{\aj}, \emph{120}, 6, 3139.

\bibitem[\protect\astroncite{\emph{{Cartwright} and
  {Whitworth}}}{2004}]{cartwright2004}
{Cartwright} A. and {Whitworth} A.~P. (2004) \emph{\mnras}, \emph{348}, 2, 589.

\bibitem[\protect\astroncite{\emph{{Chen} et~al.}}{2020}]{chen2020}
{Chen} B. et~al. (2020) \emph{\aap}, \emph{643}, A114.

\bibitem[\protect\astroncite{\emph{{Chen} et~al.}}{2013}]{chen2013}
{Chen} X. et~al. (2013) \emph{\apj}, \emph{768}, 110.

\bibitem[\protect\astroncite{\emph{{Clarke} and {Pringle}}}{1993}]{clarke1993}
{Clarke} C.~J. and {Pringle} J.~E. (1993) \emph{\mnras}, \emph{261}, 1, 190.

\bibitem[\protect\astroncite{\emph{{Comer{\'o}n} et~al.}}{2002}]{comeron2002}
{Comer{\'o}n} F. et~al. (2002) \emph{\aap}, \emph{389}, 874.

\bibitem[\protect\astroncite{\emph{{Cottaar} et~al.}}{2012}]{cott12}
{Cottaar} M. et~al. (2012) \emph{\aap}, \emph{539}, A5.

\bibitem[\protect\astroncite{\emph{{Crundall} et~al.}}{2019}]{crundall2019}
{Crundall} T.~D. et~al. (2019) \emph{\mnras}, \emph{489}, 3, 3625.

\bibitem[\protect\astroncite{\emph{{Da Rio} et~al.}}{2016}]{da-rio2016}
{Da Rio} N. et~al. (2016) \emph{\apj}, \emph{818}, 59.

\bibitem[\protect\astroncite{\emph{{Da Rio} et~al.}}{2017}]{dario2017}
{Da Rio} N. et~al. (2017) \emph{\apj}, \emph{845}, 2, 105.

\bibitem[\protect\astroncite{\emph{{Dalton} et~al.}}{2018}]{dalton2018}
{Dalton} G. et~al. (2018) in: \emph{Ground-based and Airborne Instrumentation
  for Astronomy VII}, vol. 10702 of \emph{Society of Photo-Optical
  Instrumentation Engineers (SPIE) Conference Series}, (edited by C.~J.
  {Evans}, L.~{Simard}, and H.~{Takami}), p. 107021B.

\bibitem[\protect\astroncite{\emph{{Damiani}}}{2018}]{damiani2018}
{Damiani} F. (2018) \emph{\aap}, \emph{615}, A148.

\bibitem[\protect\astroncite{\emph{{Damiani} et~al.}}{2014}]{damiani2014}
{Damiani} F. et~al. (2014) \emph{\aap}, \emph{566}, A50.

\bibitem[\protect\astroncite{\emph{{Damiani} et~al.}}{2019}]{damiani2019}
{Damiani} F. et~al. (2019) \emph{\aap}, \emph{623}, A112.

\bibitem[\protect\astroncite{\emph{{David} et~al.}}{2019}]{david2019}
{David} T.~J. et~al. (2019) \emph{\apj}, \emph{872}, 2, 161.

\bibitem[\protect\astroncite{\emph{{de Bruijne}}}{1999}]{debruijne1999}
{de Bruijne} J. H.~J. (1999) \emph{\mnras}, \emph{310}, 3, 585.

\bibitem[\protect\astroncite{\emph{{De Furio} et~al.}}{2019}]{de-furio2019}
{De Furio} M. et~al. (2019) \emph{\apj}, \emph{886}, 2, 95.

\bibitem[\protect\astroncite{\emph{{De Furio} et~al.}}{2021}]{de-furio2021}
{De Furio} M. et~al. (2021) in: \emph{American Astronomical Society Meeting
  Abstracts}, vol.~53 of \emph{American Astronomical Society Meeting
  Abstracts}, p. 414.02.

\bibitem[\protect\astroncite{\emph{{de Jong} et~al.}}{2019}]{dejong2019}
{de Jong} R.~S. et~al. (2019) \emph{The Messenger}, \emph{175}, 3.

\bibitem[\protect\astroncite{\emph{{de Juan Ovelar}
  et~al.}}{2012}]{dejuanoverlar2012}
{de Juan Ovelar} M. et~al. (2012) \emph{\aap}, \emph{546}, L1.

\bibitem[\protect\astroncite{\emph{{de Zeeuw} et~al.}}{1999}]{dezeeuw1999}
{de Zeeuw} P.~T. et~al. (1999) \emph{\aj}, \emph{117}, 1, 354.

\bibitem[\protect\astroncite{\emph{{Duch{\^e}ne} and
  {Kraus}}}{2013}]{duchene2013}
{Duch{\^e}ne} G. and {Kraus} A. (2013) \emph{\araa}, \emph{51}, 269.

\bibitem[\protect\astroncite{\emph{{Duch{\^e}ne} et~al.}}{2018}]{duchene2018}
{Duch{\^e}ne} G. et~al. (2018) \emph{\mnras}, \emph{478}, 1825.

\bibitem[\protect\astroncite{\emph{{Ducourant} et~al.}}{2014}]{Ducourant2014}
{Ducourant} C. et~al. (2014) \emph{\aap}, \emph{563}, A121.

\bibitem[\protect\astroncite{\emph{{Dzib} et~al.}}{2017}]{dzib2017}
{Dzib} S.~A. et~al. (2017) \emph{\apj}, \emph{834}, 2, 139.

\bibitem[\protect\astroncite{\emph{{Eisner} et~al.}}{2018}]{eisner2018}
{Eisner} J.~A. et~al. (2018) \emph{\apj}, \emph{860}, 1, 77.

\bibitem[\protect\astroncite{\emph{{Ekstr{\"o}m} et~al.}}{2012}]{ekstrom2012}
{Ekstr{\"o}m} S. et~al. (2012) \emph{\aap}, \emph{537}, A146.

\bibitem[\protect\astroncite{\emph{{Elmegreen} and
  {Lada}}}{1977}]{elmegreen1977}
{Elmegreen} B.~G. and {Lada} C.~J. (1977) \emph{\apj}, \emph{214}, 725.

\bibitem[\protect\astroncite{\emph{{Fedele} et~al.}}{2010}]{Fedele2010}
{Fedele} D. et~al. (2010) \emph{\aap}, \emph{510}, A72.

\bibitem[\protect\astroncite{\emph{{Feiden}}}{2016}]{feiden2016}
{Feiden} G.~A. (2016) \emph{\aap}, \emph{593}, A99.

\bibitem[\protect\astroncite{\emph{{Feigelson} et~al.}}{2013}]{feigelson2013}
{Feigelson} E.~D. et~al. (2013) \emph{\apjs}, \emph{209}, 2, 26.

\bibitem[\protect\astroncite{\emph{{F{\H{u}}r{\'e}sz}
  et~al.}}{2008}]{furesz2008}
{F{\H{u}}r{\'e}sz} G. et~al. (2008) \emph{\apj}, \emph{676}, 2, 1109.

\bibitem[\protect\astroncite{\emph{{Fischer} et~al.}}{2017}]{fischer2017}
{Fischer} W.~J. et~al. (2017) \emph{\apj}, \emph{840}, 69.

\bibitem[\protect\astroncite{\emph{{Fitzgerald} et~al.}}{1990}]{fitzgerald1990}
{Fitzgerald} M.~P. et~al. (1990) \emph{\pasp}, \emph{102}, 865.

\bibitem[\protect\astroncite{\emph{{Galli} et~al.}}{2019}]{galli2019}
{Galli} P.~A.~B. et~al. (2019) \emph{\aap}, \emph{630}, A137.

\bibitem[\protect\astroncite{\emph{{Garmany}}}{1994}]{garmany1994}
{Garmany} C.~D. (1994) \emph{\pasp}, \emph{106}, 25.

\bibitem[\protect\astroncite{\emph{{Garmany} and
  {Stencel}}}{1992}]{garmany1992}
{Garmany} C.~D. and {Stencel} R.~E. (1992) \emph{\aaps}, \emph{94}, 211.

\bibitem[\protect\astroncite{\emph{{Gieles} and {Portegies
  Zwart}}}{2011}]{gieles2011}
{Gieles} M. and {Portegies Zwart} S.~F. (2011) \emph{\mnras}, \emph{410}, 1,
  L6.

\bibitem[\protect\astroncite{\emph{{Gieles} et~al.}}{2010}]{gieles2010}
{Gieles} M. et~al. (2010) \emph{\mnras}, \emph{402}, 3, 1750.

\bibitem[\protect\astroncite{\emph{{Gonz{\'a}lez} and
  {Alfaro}}}{2017}]{gonzalez2017}
{Gonz{\'a}lez} M. and {Alfaro} E.~J. (2017) \emph{\mnras}, \emph{465}, 2, 1889.

\bibitem[\protect\astroncite{\emph{{Goodwin}}}{1997}]{goodwin1997}
{Goodwin} S.~P. (1997) \emph{\mnras}, \emph{286}, 3, 669.

\bibitem[\protect\astroncite{\emph{{Goodwin} and
  {Bastian}}}{2006}]{goodwin2006}
{Goodwin} S.~P. and {Bastian} N. (2006) \emph{\mnras}, \emph{373}, 2, 752.

\bibitem[\protect\astroncite{\emph{{Goodwin} and {Kroupa}}}{2005}]{goodwin2005}
{Goodwin} S.~P. and {Kroupa} P. (2005) \emph{\aap}, \emph{439}, 2, 565.

\bibitem[\protect\astroncite{\emph{{Goodwin} and
  {Whitworth}}}{2004}]{goodwin2004}
{Goodwin} S.~P. and {Whitworth} A.~P. (2004) \emph{\aap}, \emph{413}, 929.

\bibitem[\protect\astroncite{\emph{{Goodwin} et~al.}}{2007}]{goodwin2007}
{Goodwin} S.~P. et~al. (2007) in: \emph{Protostars and Planets V}, (edited by
  B.~{Reipurth}, D.~{Jewitt}, and K.~{Keil}), p. 133.

\bibitem[\protect\astroncite{\emph{{Gouliermis}}}{2018}]{gouliermis2018}
{Gouliermis} D.~A. (2018) \emph{\pasp}, \emph{130}, 989, 072001.

\bibitem[\protect\astroncite{\emph{{Green} et~al.}}{2019}]{Green2019}
{Green} G.~M. et~al. (2019) \emph{\apj}, \emph{887}, 1, 93.

\bibitem[\protect\astroncite{\emph{{Griffiths} et~al.}}{2018}]{griffiths2018}
{Griffiths} D.~W. et~al. (2018) \emph{\mnras}, \emph{476}, 2, 2493.

\bibitem[\protect\astroncite{\emph{{Gro{\ss}schedl}
  et~al.}}{2021}]{grossschedl2021}
{Gro{\ss}schedl} J.~E. et~al. (2021) \emph{\aap}, \emph{647}, A91.

\bibitem[\protect\astroncite{\emph{{Guarcello} et~al.}}{2016}]{guarcello2016}
{Guarcello} M.~G. et~al. (2016) \emph{arXiv e-prints}, arXiv:1605.01773.

\bibitem[\protect\astroncite{\emph{{Gully-Santiago} et~al.}}{2017}]{gully2017}
{Gully-Santiago} M.~A. et~al. (2017) \emph{\apj}, \emph{836}, 2, 200.

\bibitem[\protect\astroncite{\emph{{Gutermuth} et~al.}}{2009}]{gutermuth2009}
{Gutermuth} R.~A. et~al. (2009) \emph{\apjs}, \emph{184}, 1, 18.

\bibitem[\protect\astroncite{\emph{{Haworth} and
  {Harries}}}{2012}]{haworth2012}
{Haworth} T.~J. and {Harries} T.~J. (2012) \emph{\mnras}, \emph{420}, 1, 562.

\bibitem[\protect\astroncite{\emph{{Heggie}}}{1975}]{heggie1975}
{Heggie} D.~C. (1975) \emph{\mnras}, \emph{173}, 729.

\bibitem[\protect\astroncite{\emph{{Henney} et~al.}}{2002}]{henney2002}
{Henney} W.~J. et~al. (2002) \emph{\apj}, \emph{566}, 1, 315.

\bibitem[\protect\astroncite{\emph{{Herbig}}}{1962}]{herbig1962}
{Herbig} G.~H. (1962) \emph{Advances in Astronomy and Astrophysics}, \emph{1},
  47.

\bibitem[\protect\astroncite{\emph{{Hillenbrand}
  et~al.}}{2008}]{hillenbrand2008}
{Hillenbrand} L.~A. et~al. (2008) in: \emph{14th Cambridge Workshop on Cool
  Stars, Stellar Systems, and the Sun}, vol. 384 of \emph{Astronomical Society
  of the Pacific Conference Series}, (edited by G.~{van Belle}), p. 200.

\bibitem[\protect\astroncite{\emph{{Hills}}}{1975}]{hills1975}
{Hills} J.~G. (1975) \emph{\aj}, \emph{80}, 809.

\bibitem[\protect\astroncite{\emph{{Hills}}}{1980}]{hills1980}
{Hills} J.~G. (1980) \emph{\apj}, \emph{235}, 986.

\bibitem[\protect\astroncite{\emph{{Hoogerwerf} and
  {Aguilar}}}{1999}]{hoogerwerf1999}
{Hoogerwerf} R. and {Aguilar} L.~A. (1999) \emph{\mnras}, \emph{306}, 2, 394.

\bibitem[\protect\astroncite{\emph{{Humphreys}}}{1978}]{humphreys1978}
{Humphreys} R.~M. (1978) \emph{\apjs}, \emph{38}, 309.

\bibitem[\protect\astroncite{\emph{{Ivanov}}}{1987}]{ivanov1987}
{Ivanov} G.~R. (1987) \emph{\apss}, \emph{136}, 1, 113.

\bibitem[\protect\astroncite{\emph{{Jackson} et~al.}}{2015}]{jack15}
{Jackson} R.~J. et~al. (2015) \emph{\aap}, \emph{580}.

\bibitem[\protect\astroncite{\emph{{Jackson} et~al.}}{2020}]{jack20}
{Jackson} R.~J. et~al. (2020) \emph{\mnras}, \emph{496}, 4, 4701.

\bibitem[\protect\astroncite{\emph{{Jeffries}}}{2014}]{Jeffries2014b}
{Jeffries} R.~D. (2014) in: \emph{EAS Publications Series}, vol.~65 of
  \emph{EAS Publications Series}, pp. 289--325.

\bibitem[\protect\astroncite{\emph{{Jeffries} et~al.}}{2011}]{jeffries2011}
{Jeffries} R.~D. et~al. (2011) \emph{\mnras}, \emph{418}, 3, 1948.

\bibitem[\protect\astroncite{\emph{{Jeffries} et~al.}}{2014}]{jeffries2014}
{Jeffries} R.~D. et~al. (2014) \emph{\aap}, \emph{563}, A94.

\bibitem[\protect\astroncite{\emph{{Jeffries} et~al.}}{2017}]{jeffries2017}
{Jeffries} R.~D. et~al. (2017) \emph{\mnras}, \emph{464}, 2, 1456.

\bibitem[\protect\astroncite{\emph{{Jerabkova} et~al.}}{2019}]{jerabkova2019}
{Jerabkova} T. et~al. (2019) \emph{\mnras}, \emph{489}, 3, 4418.

\bibitem[\protect\astroncite{\emph{{Johnson} and {Morgan}}}{1953}]{johnson1953}
{Johnson} H.~L. and {Morgan} W.~W. (1953) \emph{\apj}, \emph{117}, 313.

\bibitem[\protect\astroncite{\emph{{Joy}}}{1945}]{joy1945}
{Joy} A.~H. (1945) \emph{\apj}, \emph{102}, 168.

\bibitem[\protect\astroncite{\emph{{Kalas} et~al.}}{2015}]{kalas2015}
{Kalas} P.~G. et~al. (2015) \emph{\apj}, \emph{814}, 1, 32.

\bibitem[\protect\astroncite{\emph{{Kerr} et~al.}}{2021}]{kerr2021}
{Kerr} R. M.~P. et~al. (2021) \emph{\apj}, \emph{917}, 1, 23.

\bibitem[\protect\astroncite{\emph{{Kessel-Deynet} and
  {Burkert}}}{2003}]{kessel2003}
{Kessel-Deynet} O. and {Burkert} A. (2003) \emph{\mnras}, \emph{338}, 3, 545.

\bibitem[\protect\astroncite{\emph{{King} et~al.}}{2012}]{king2012}
{King} R.~R. et~al. (2012) \emph{\mnras}, \emph{427}, 3, 2636.

\bibitem[\protect\astroncite{\emph{{Kollmeier} et~al.}}{2017}]{Kollmeier2017}
{Kollmeier} J.~A. et~al. (2017) \emph{arXiv e-prints}, arXiv:1711.03234.

\bibitem[\protect\astroncite{\emph{{K{\"o}nyves} et~al.}}{2010}]{konyves2010}
{K{\"o}nyves} V. et~al. (2010) \emph{\aap}, \emph{518}, L106.

\bibitem[\protect\astroncite{\emph{{K{\"o}nyves} et~al.}}{2020}]{konyves2020}
{K{\"o}nyves} V. et~al. (2020) \emph{\aap}, \emph{635}, A34.

\bibitem[\protect\astroncite{\emph{{Kos} et~al.}}{2019}]{kos2019}
{Kos} J. et~al. (2019) \emph{\aap}, \emph{631}, A166.

\bibitem[\protect\astroncite{\emph{{Kounkel}}}{2020}]{kounkel2020}
{Kounkel} M. (2020) \emph{\apj}, \emph{902}, 2, 122.

\bibitem[\protect\astroncite{\emph{{Kounkel} and {Covey}}}{2019}]{kounkel2019}
{Kounkel} M. and {Covey} K. (2019) \emph{\aj}, \emph{158}, 3, 122.

\bibitem[\protect\astroncite{\emph{{Kounkel} et~al.}}{2016}]{kounkel2016}
{Kounkel} M. et~al. (2016) \emph{\apj}, \emph{821}, 52.

\bibitem[\protect\astroncite{\emph{{Kounkel} et~al.}}{2018}]{kounkel2018}
{Kounkel} M. et~al. (2018) \emph{\aj}, \emph{156}, 3, 84.

\bibitem[\protect\astroncite{\emph{{Kouwenhoven}
  et~al.}}{2007}]{kouwenhoven2007}
{Kouwenhoven} M.~B.~N. et~al. (2007) \emph{\aap}, \emph{474}, 1, 77.

\bibitem[\protect\astroncite{\emph{{Kraus} et~al.}}{2011}]{kraus2011}
{Kraus} A.~L. et~al. (2011) \emph{\apj}, \emph{731}, 8.

\bibitem[\protect\astroncite{\emph{{Kraus} et~al.}}{2015}]{kraus2015}
{Kraus} A.~L. et~al. (2015) \emph{\apj}, \emph{807}, 1, 3.

\bibitem[\protect\astroncite{\emph{{Krause} et~al.}}{2018}]{krause2018}
{Krause} M. G.~H. et~al. (2018) \emph{\aap}, \emph{619}, A120.

\bibitem[\protect\astroncite{\emph{{Krolikowski}
  et~al.}}{2021}]{krolikowski2021}
{Krolikowski} D.~M. et~al. (2021) \emph{\aj}, \emph{162}, 3, 110.

\bibitem[\protect\astroncite{\emph{{Kroupa}}}{1995}]{kroupa1995}
{Kroupa} P. (1995) \emph{\mnras}, \emph{277}, 1491.

\bibitem[\protect\astroncite{\emph{{Kroupa}}}{2001}]{kroupa2001}
{Kroupa} P. (2001) \emph{\mnras}, \emph{322}, 231.

\bibitem[\protect\astroncite{\emph{{Kroupa}}}{2011}]{kroupa2011}
{Kroupa} P. (2011) in: \emph{Stellar Clusters \& Associations: A RIA Workshop
  on Gaia}, pp. 17--27.

\bibitem[\protect\astroncite{\emph{{Kruijssen}}}{2012}]{kruijssen2012}
{Kruijssen} J.~M.~D. (2012) \emph{\mnras}, \emph{426}, 4, 3008.

\bibitem[\protect\astroncite{\emph{{Kuhn} et~al.}}{2014}]{kuhn2014}
{Kuhn} M.~A. et~al. (2014) \emph{\apj}, \emph{787}, 107.

\bibitem[\protect\astroncite{\emph{{Kuhn} et~al.}}{2019}]{kuhn2019}
{Kuhn} M.~A. et~al. (2019) \emph{\apj}, \emph{870}, 1, 32.

\bibitem[\protect\astroncite{\emph{{Kuhn} et~al.}}{2020}]{kuhn2020}
{Kuhn} M.~A. et~al. (2020) \emph{\apj}, \emph{899}, 2, 128.

\bibitem[\protect\astroncite{\emph{{Kuhn}
  et~al.}}{2021{\natexlab{a}}}]{Kuhn2021b}
{Kuhn} M.~A. et~al. (2021{\natexlab{a}}) \emph{\aap}, \emph{651}, L10.

\bibitem[\protect\astroncite{\emph{{Kuhn}
  et~al.}}{2021{\natexlab{b}}}]{Kuhn2021a}
{Kuhn} M.~A. et~al. (2021{\natexlab{b}}) \emph{\apjs}, \emph{254}, 2, 33.

\bibitem[\protect\astroncite{\emph{{Lada} and {Lada}}}{2003}]{lada2003}
{Lada} C.~J. and {Lada} E.~A. (2003) \emph{\araa}, \emph{41}, 57.

\bibitem[\protect\astroncite{\emph{{Lallement} et~al.}}{2019}]{Lallement2019}
{Lallement} R. et~al. (2019) \emph{\aap}, \emph{625}, A135.

\bibitem[\protect\astroncite{\emph{{Larson}}}{1981}]{larson1981}
{Larson} R.~B. (1981) \emph{\mnras}, \emph{194}, 809.

\bibitem[\protect\astroncite{\emph{{Laughlin} and
  {Adams}}}{1998}]{laughlin1998}
{Laughlin} G. and {Adams} F.~C. (1998) \emph{\apjl}, \emph{508}, 2, L171.

\bibitem[\protect\astroncite{\emph{{Lesh}}}{1968}]{lesh1968}
{Lesh} J.~R. (1968) \emph{\apj}, \emph{152}, 905.

\bibitem[\protect\astroncite{\emph{{Liu} et~al.}}{2021}]{liu2021}
{Liu} J. et~al. (2021) \emph{\apjs}, \emph{254}, 1, 20.

\bibitem[\protect\astroncite{\emph{{L{\'o}pez-Valdivia}
  et~al.}}{2021}]{lopez-valdivia2021}
{L{\'o}pez-Valdivia} R. et~al. (2021) \emph{\apj}, \emph{921}, 1, 53.

\bibitem[\protect\astroncite{\emph{{Lucke} and {Hodge}}}{1970}]{lucke1970}
{Lucke} P.~B. and {Hodge} P.~W. (1970) \emph{\aj}, \emph{75}, 171.

\bibitem[\protect\astroncite{\emph{{Makarov}}}{2007}]{makarov2007}
{Makarov} V.~V. (2007) \emph{\apjs}, \emph{169}, 1, 105.

\bibitem[\protect\astroncite{\emph{{Mamajek} et~al.}}{2002}]{Mamajek2002}
{Mamajek} E.~E. et~al. (2002) \emph{\aj}, \emph{124}, 3, 1670.

\bibitem[\protect\astroncite{\emph{{Manara} et~al.}}{2018}]{manara2018}
{Manara} C.~F. et~al. (2018) \emph{\aap}, \emph{615}, L1.

\bibitem[\protect\astroncite{\emph{{Marigo} et~al.}}{2017}]{Marigo2017}
{Marigo} P. et~al. (2017) \emph{\apj}, \emph{835}, 1, 77.

\bibitem[\protect\astroncite{\emph{{Mart{\'\i}n}}}{1998}]{Martin1998}
{Mart{\'\i}n} E.~L. (1998) \emph{\aj}, \emph{115}, 1, 351.

\bibitem[\protect\astroncite{\emph{{Marton} et~al.}}{2016}]{marton2016}
{Marton} G. et~al. (2016) \emph{\mnras}, \emph{458}, 3479.

\bibitem[\protect\astroncite{\emph{{Massey} et~al.}}{1995}]{massey1995}
{Massey} P. et~al. (1995) \emph{\apj}, \emph{454}, 151.

\bibitem[\protect\astroncite{\emph{{McBride} et~al.}}{2021}]{mcbride2021}
{McBride} A. et~al. (2021) \emph{\aj}, \emph{162}, 6, 282.

\bibitem[\protect\astroncite{\emph{{Megeath} et~al.}}{2016}]{megeath2016}
{Megeath} S.~T. et~al. (2016) \emph{\aj}, \emph{151}, 5.

\bibitem[\protect\astroncite{\emph{{Mel'nik} and {Dambis}}}{2017}]{melnik2017}
{Mel'nik} A.~M. and {Dambis} A.~K. (2017) \emph{\mnras}, \emph{472}, 4, 3887.

\bibitem[\protect\astroncite{\emph{{Melnik} and {Dambis}}}{2020}]{melnik2020}
{Melnik} A.~M. and {Dambis} A.~K. (2020) \emph{\mnras}, \emph{493}, 2, 2339.

\bibitem[\protect\astroncite{\emph{{Mel'Nik} and {Efremov}}}{1995}]{melnik1995}
{Mel'Nik} A.~M. and {Efremov} Y.~N. (1995) \emph{Astronomy Letters}, \emph{21},
  1, 10.

\bibitem[\protect\astroncite{\emph{{Merloni} et~al.}}{2020}]{merloni2020}
{Merloni} A. et~al. (2020) \emph{Nature Astronomy}, \emph{4}, 634.

\bibitem[\protect\astroncite{\emph{{Miller} and {Scalo}}}{1978}]{miller1978}
{Miller} G.~E. and {Scalo} J.~M. (1978) \emph{\pasp}, \emph{90}, 506.

\bibitem[\protect\astroncite{\emph{{Miret-Roig} et~al.}}{2018}]{miret-roig2018}
{Miret-Roig} N. et~al. (2018) \emph{\aap}, \emph{615}, A51.

\bibitem[\protect\astroncite{\emph{{Moeckel} and {Clarke}}}{2011}]{moeckel2011}
{Moeckel} N. and {Clarke} C.~J. (2011) \emph{\mnras}, \emph{415}, 2, 1179.

\bibitem[\protect\astroncite{\emph{{Mohr-Smith} et~al.}}{2015}]{mohr-smith2015}
{Mohr-Smith} M. et~al. (2015) \emph{\mnras}, \emph{450}, 4, 3855.

\bibitem[\protect\astroncite{\emph{{Mohr-Smith} et~al.}}{2017}]{mohr-smith2017}
{Mohr-Smith} M. et~al. (2017) \emph{\mnras}, \emph{465}, 2, 1807.

\bibitem[\protect\astroncite{\emph{{Morgan} et~al.}}{1953}]{morgan1953}
{Morgan} W.~W. et~al. (1953) \emph{\apj}, \emph{118}, 318.

\bibitem[\protect\astroncite{\emph{{Nicholson} et~al.}}{2019}]{nicholson2019}
{Nicholson} R.~B. et~al. (2019) \emph{\mnras}, \emph{485}, 4, 4893.

\bibitem[\protect\astroncite{\emph{{Olney} et~al.}}{2020}]{olney2020}
{Olney} R. et~al. (2020) \emph{\aj}, \emph{159}, 4, 182.

\bibitem[\protect\astroncite{\emph{{Parker}}}{2020}]{parker2020}
{Parker} R.~J. (2020) \emph{Royal Society Open Science}, \emph{7}, 11, 201271.

\bibitem[\protect\astroncite{\emph{{Parker} and {Goodwin}}}{2012}]{parker2012b}
{Parker} R.~J. and {Goodwin} S.~P. (2012) \emph{\mnras}, \emph{424}, 1, 272.

\bibitem[\protect\astroncite{\emph{{Parker} and {Wright}}}{2016}]{parker2016}
{Parker} R.~J. and {Wright} N.~J. (2016) \emph{\mnras}, \emph{457}, 4, 3430.

\bibitem[\protect\astroncite{\emph{{Parker} et~al.}}{2014}]{parker2014}
{Parker} R.~J. et~al. (2014) \emph{\mnras}, \emph{438}, 1, 620.

\bibitem[\protect\astroncite{\emph{{Parker} et~al.}}{2021}]{parker2021}
{Parker} R.~J. et~al. (2021) \emph{\apj}, \emph{913}, 2, 95.

\bibitem[\protect\astroncite{\emph{{Pecaut} and {Mamajek}}}{2016}]{pecaut2016}
{Pecaut} M.~J. and {Mamajek} E.~E. (2016) \emph{\mnras}, \emph{461}, 1, 794.

\bibitem[\protect\astroncite{\emph{{Pfalzner}}}{2009}]{pfalzner2009}
{Pfalzner} S. (2009) \emph{\aap}, \emph{498}, 2, L37.

\bibitem[\protect\astroncite{\emph{{Poggio} et~al.}}{2018}]{poggio2018}
{Poggio} E. et~al. (2018) \emph{\mnras}, \emph{481}, 1, L21.

\bibitem[\protect\astroncite{\emph{{Portegies Zwart} and
  {J{\'\i}lkov{\'a}}}}{2015}]{portegies-zwart2015}
{Portegies Zwart} S.~F. and {J{\'\i}lkov{\'a}} L. (2015) \emph{\mnras},
  \emph{451}, 1, 144.

\bibitem[\protect\astroncite{\emph{{Portegies Zwart}
  et~al.}}{2010}]{portegieszwart2010}
{Portegies Zwart} S.~F. et~al. (2010) \emph{\araa}, \emph{48}, 431.

\bibitem[\protect\astroncite{\emph{{Predehl} et~al.}}{2021}]{predehl2021}
{Predehl} P. et~al. (2021) \emph{\aap}, \emph{647}, A1.

\bibitem[\protect\astroncite{\emph{{Preibisch} and
  {Mamajek}}}{2008}]{preibisch2008}
{Preibisch} T. and {Mamajek} E. (2008) \emph{{The Nearest OB Association:
  Scorpius-Centaurus (Sco OB2)}}, vol.~5, p. 235, Handbook of Star Forming
  Regions.

\bibitem[\protect\astroncite{\emph{{Preibisch} and
  {Zinnecker}}}{2007}]{preibisch2007}
{Preibisch} T. and {Zinnecker} H. (2007) in: \emph{Triggered Star Formation in
  a Turbulent ISM}, vol. 237, (edited by B.~G. {Elmegreen} and J.~{Palous}),
  pp. 270--277.

\bibitem[\protect\astroncite{\emph{{Preibisch} et~al.}}{1998}]{preibisch1998}
{Preibisch} T. et~al. (1998) \emph{\aap}, \emph{333}, 619.

\bibitem[\protect\astroncite{\emph{{Preibisch} et~al.}}{2002}]{preibisch2002}
{Preibisch} T. et~al. (2002) \emph{\aj}, \emph{124}, 1, 404.

\bibitem[\protect\astroncite{\emph{{Prisinzano} et~al.}}{2016}]{Prisinzano2016}
{Prisinzano} L. et~al. (2016) \emph{\aap}, \emph{589}, A70.

\bibitem[\protect\astroncite{\emph{{Quintana} and
  {Wright}}}{2021}]{quintana2021}
{Quintana} A.~L. and {Wright} N.~J. (2021) \emph{\mnras}, \emph{508}, 2, 2370.

\bibitem[\protect\astroncite{\emph{{Reggiani} et~al.}}{2011}]{reggiani2011}
{Reggiani} M. et~al. (2011) \emph{\aap}, \emph{534}, A83.

\bibitem[\protect\astroncite{\emph{{Reid} et~al.}}{2019}]{reid2019}
{Reid} M.~J. et~al. (2019) \emph{\apj}, \emph{885}, 2, 131.

\bibitem[\protect\astroncite{\emph{{Reipurth} et~al.}}{2007}]{reipurth2007}
{Reipurth} B. et~al. (2007) \emph{\aj}, \emph{134}, 2272.

\bibitem[\protect\astroncite{\emph{{Reipurth} et~al.}}{2010}]{reipurth2010}
{Reipurth} B. et~al. (2010) \emph{\apjl}, \emph{725}, L56.

\bibitem[\protect\astroncite{\emph{{Reipurth} et~al.}}{2014}]{reipurth2014}
{Reipurth} B. et~al. (2014) in: \emph{Protostars and Planets VI}, (edited by
  H.~{Beuther}, R.~S. {Klessen}, C.~P. {Dullemond}, and T.~{Henning}), p. 267.

\bibitem[\protect\astroncite{\emph{{Rizzuto} et~al.}}{2011}]{rizzuto2011}
{Rizzuto} A.~C. et~al. (2011) \emph{\mnras}, \emph{416}, 4, 3108.

\bibitem[\protect\astroncite{\emph{{Ruprecht}}}{1966}]{ruprecht1966}
{Ruprecht} J. (1966) \emph{IAU Trans.}, \emph{12}, 348.

\bibitem[\protect\astroncite{\emph{{Sacco} et~al.}}{2015}]{sacco2015}
{Sacco} G.~G. et~al. (2015) \emph{\aap}, \emph{574}, L7.

\bibitem[\protect\astroncite{\emph{{Sandford} et~al.}}{1982}]{Sandford1982}
{Sandford} M.~T. I. et~al. (1982) \emph{\apj}, \emph{260}, 183.

\bibitem[\protect\astroncite{\emph{{Scally} et~al.}}{1999}]{scally1999}
{Scally} A. et~al. (1999) \emph{\mnras}, \emph{306}, 1, 253.

\bibitem[\protect\astroncite{\emph{{Schmitt} et~al.}}{2021}]{Schmitt2021}
{Schmitt} J.~H.~M.~M. et~al. (2021) \emph{arXiv e-prints}, arXiv:2106.14549.

\bibitem[\protect\astroncite{\emph{{Scholz} et~al.}}{2009}]{scholz2009}
{Scholz} A. et~al. (2009) \emph{\apj}, \emph{702}, 1, 805.

\bibitem[\protect\astroncite{\emph{{Sherry} et~al.}}{2004}]{sherry2004}
{Sherry} W.~H. et~al. (2004) \emph{\aj}, \emph{128}, 5, 2316.

\bibitem[\protect\astroncite{\emph{{Soderblom}}}{2010}]{soderblom2010}
{Soderblom} D.~R. (2010) \emph{\araa}, \emph{48}, 581.

\bibitem[\protect\astroncite{\emph{{Soderblom} et~al.}}{2014}]{soderblom2014}
{Soderblom} D.~R. et~al. (2014) in: \emph{Protostars and Planets VI}, (edited
  by H.~{Beuther}, R.~S. {Klessen}, C.~P. {Dullemond}, and T.~{Henning}), p.
  219.

\bibitem[\protect\astroncite{\emph{{Somers} et~al.}}{2020}]{somers2020}
{Somers} G. et~al. (2020) \emph{\apj}, \emph{891}, 1, 29.

\bibitem[\protect\astroncite{\emph{{Squicciarini}
  et~al.}}{2021}]{squicciarini2021}
{Squicciarini} V. et~al. (2021) \emph{\mnras}.

\bibitem[\protect\astroncite{\emph{{Sterzik} et~al.}}{1995}]{sterzik1995}
{Sterzik} M.~F. et~al. (1995) \emph{\aap}, \emph{297}, 418.

\bibitem[\protect\astroncite{\emph{{Tobin} et~al.}}{2009}]{tobin2009}
{Tobin} J.~J. et~al. (2009) \emph{\apj}, \emph{697}, 2, 1103.

\bibitem[\protect\astroncite{\emph{{Tobin} et~al.}}{2019}]{tobin2019}
{Tobin} J.~J. et~al. (2019) \emph{\apj}, \emph{886}, 1, 6.

\bibitem[\protect\astroncite{\emph{{Tobin} et~al.}}{2022}]{tobin2022}
{Tobin} J.~J. et~al. (2022) \emph{\apj}, \emph{925}, 1, 39.

\bibitem[\protect\astroncite{\emph{{Torres} et~al.}}{2008}]{torres2008}
{Torres} C.~A.~O. et~al. (2008) \emph{{Young Nearby Loose Associations}},
  vol.~5, p. 757, Handbook of Star Forming Regions.

\bibitem[\protect\astroncite{\emph{{T{\'o}th} et~al.}}{2014}]{toth2014}
{T{\'o}th} L.~V. et~al. (2014) \emph{\pasj}, \emph{66}, 1, 17.

\bibitem[\protect\astroncite{\emph{{Vincke} et~al.}}{2015}]{vincke2015}
{Vincke} K. et~al. (2015) \emph{\aap}, \emph{577}, A115.

\bibitem[\protect\astroncite{\emph{{Walter} et~al.}}{1994}]{walter1994}
{Walter} F.~M. et~al. (1994) \emph{\aj}, \emph{107}, 692.

\bibitem[\protect\astroncite{\emph{{Walter} et~al.}}{2000}]{walter2000}
{Walter} F.~M. et~al. (2000) in: \emph{Protostars and Planets IV}, (edited by
  V.~{Mannings}, A.~P. {Boss}, and S.~S. {Russell}), p. 273.

\bibitem[\protect\astroncite{\emph{{Ward} and {Kruijssen}}}{2018}]{ward2018}
{Ward} J.~L. and {Kruijssen} J.~M.~D. (2018) \emph{\mnras}.

\bibitem[\protect\astroncite{\emph{{Ward} et~al.}}{2020}]{ward2020}
{Ward} J.~L. et~al. (2020) \emph{\mnras}, \emph{495}, 1, 663.

\bibitem[\protect\astroncite{\emph{{White} and {Basri}}}{2003}]{white2003}
{White} R.~J. and {Basri} G. (2003) \emph{\apj}, \emph{582}, 1109.

\bibitem[\protect\astroncite{\emph{{Wilking} et~al.}}{2005}]{wilking2005}
{Wilking} B.~A. et~al. (2005) \emph{\aj}, \emph{130}, 1733.

\bibitem[\protect\astroncite{\emph{{Winston} et~al.}}{2020}]{winston2020}
{Winston} E. et~al. (2020) \emph{\aj}, \emph{160}, 2, 68.

\bibitem[\protect\astroncite{\emph{{Winter} et~al.}}{2018}]{winter2018b}
{Winter} A.~J. et~al. (2018) \emph{\mnras}, \emph{478}, 2, 2700.

\bibitem[\protect\astroncite{\emph{{Winter}
  et~al.}}{2019{\natexlab{a}}}]{winter2019a}
{Winter} A.~J. et~al. (2019{\natexlab{a}}) \emph{\mnras}, \emph{490}, 4, 5478.

\bibitem[\protect\astroncite{\emph{{Winter}
  et~al.}}{2019{\natexlab{b}}}]{winter2019}
{Winter} A.~J. et~al. (2019{\natexlab{b}}) \emph{\mnras}, \emph{485}, 2, 1489.

\bibitem[\protect\astroncite{\emph{{Wright}}}{2020}]{wright2020}
{Wright} N.~J. (2020) \emph{\nar}, \emph{90}, 101549.

\bibitem[\protect\astroncite{\emph{{Wright} and {Mamajek}}}{2018}]{wright2018}
{Wright} N.~J. and {Mamajek} E.~E. (2018) \emph{\mnras}, \emph{476}, 381.

\bibitem[\protect\astroncite{\emph{{Wright} et~al.}}{2010}]{wright2010}
{Wright} N.~J. et~al. (2010) \emph{\apj}, \emph{713}, 2, 871.

\bibitem[\protect\astroncite{\emph{{Wright} et~al.}}{2012}]{wright2012}
{Wright} N.~J. et~al. (2012) \emph{\apjl}, \emph{746}, 2, L21.

\bibitem[\protect\astroncite{\emph{{Wright} et~al.}}{2014}]{wright2014}
{Wright} N.~J. et~al. (2014) \emph{\mnras}, \emph{438}, 1, 639.

\bibitem[\protect\astroncite{\emph{{Wright} et~al.}}{2015}]{wright2015}
{Wright} N.~J. et~al. (2015) \emph{\mnras}, \emph{449}, 1, 741.

\bibitem[\protect\astroncite{\emph{{Wright} et~al.}}{2016}]{wright2016}
{Wright} N.~J. et~al. (2016) \emph{\mnras}, \emph{460}, 3, 2593.

\bibitem[\protect\astroncite{\emph{{Wright} et~al.}}{2019}]{Wright2019}
{Wright} N.~J. et~al. (2019) \emph{\mnras}, \emph{486}, 2, 2477.

\bibitem[\protect\astroncite{\emph{{Zari} et~al.}}{2017}]{zari2017}
{Zari} E. et~al. (2017) \emph{\aap}, \emph{608}, A148.

\bibitem[\protect\astroncite{\emph{{Zari} et~al.}}{2018}]{zari2018}
{Zari} E. et~al. (2018) \emph{\aap}, \emph{620}, A172.

\bibitem[\protect\astroncite{\emph{{Zari} et~al.}}{2019}]{zari2019}
{Zari} E. et~al. (2019) \emph{\aap}, \emph{628}, A123.

\bibitem[\protect\astroncite{\emph{{Zari} et~al.}}{2021}]{zari2021}
{Zari} E. et~al. (2021) \emph{\aap}, \emph{650}, A112.

\bibitem[\protect\astroncite{\emph{{Zinnecker} and
  {Yorke}}}{2007}]{zinnecker2007}
{Zinnecker} H. and {Yorke} H.~W. (2007) \emph{\araa}, \emph{45}, 1, 481.

\bibitem[\protect\astroncite{\emph{{Zucker} et~al.}}{2020}]{Zucker2020}
{Zucker} C. et~al. (2020) \emph{\aap}, \emph{633}, A51.

\end{thebibliography}

\end{document}